\documentclass[12pt]{article}
\usepackage{epsfig,amssymb,euscript,bbold}
\usepackage{amsfonts}
\usepackage{eufrak}
\usepackage{bm}
\usepackage{amsmath}

\newcommand{\sss}{\sqrt{1-\lambda^3\rho^2}}
\def\be#1\ee{\begin{equation}#1\end{equation}}
\newcommand{\bea}{\begin{eqnarray}}
\newcommand{\eea}{\end{eqnarray}}
\newcommand{\ba}{\begin{array}}

\newcommand{\ea}{\end{array}}

\def\bbox{{\,\lower0.9pt\vbox{\hrule \hbox{\vrule height 0.2 cm
\hskip 0.2 cm \vrule height 0.2 cm}\hrule}\,}}
\newcommand{\dsl}{\pa \kern-0.5em /}

\newcommand{\nn}{\nonumber \\}

\def\l{\lambda}
\def\r{\rho}
\def\a{\alpha}
\def\e{\epsilon}
\def\sss{\sqrt{1-\l^3\r^2}}

\def\ds{\raise.15ex\hbox{/}\kern-.57em\partial}
\def\Ds{\,\raise.15ex\hbox{/}\mkern-13.5mu D}
\newcommand{\dd}{\mathrm{d}}
\newcommand{\ee}{\mathrm{e}}
\newcommand{\ii}{\mathrm{i}}

\newcommand{\der}{\partial}

\newcommand{\bbR}{\mathbb{R}}

\DeclareMathOperator{\SU}{\mathit{SU}}
\DeclareMathOperator{\SO}{\mathit{SO}}
\DeclareMathOperator{\Symp}{\mathit{Sp}}
\DeclareMathOperator{\Spin}{\mathit{Spin}}

\DeclareMathOperator{\re}{Re}
\DeclareMathOperator{\im}{Im}

\DeclareMathOperator{\vol}{vol}
\DeclareMathOperator{\hvol}{\widehat{vol}}

\begin{document}

\makeatletter
\renewcommand{\theequation}{\thesection.\arabic{equation}}
\@addtoreset{equation}{section}
\makeatother

\baselineskip 18pt

\begin{titlepage}

\vfill

\begin{flushright}
Imperial/TP/2006/OC/01\\
\end{flushright}

\vfill

\begin{center}
   \baselineskip=16pt
   {\Large\bf AdS spacetimes from wrapped M5 branes}
   \vskip 2cm
      Jerome P. Gauntlett, Ois\'{\i}n A. P. Mac Conamhna, Toni Mateos \\
      and Daniel Waldram
   \vskip .6cm
      \begin{small}
      \textit{Theoretical Physics Group, Blackett Laboratory, \\
        Imperial College, London SW7 2AZ, U.K.}
        \end{small}\\*[.6cm]
      \begin{small}
      \textit{The Institute for Mathematical Sciences, \\
        Imperial College, London SW7 2PE, U.K.}
        \end{small}
   \end{center}

\vfill

\begin{center}
\textbf{Abstract}
\end{center}

\begin{quote}

We derive a complete geometrical characterisation of a large class of
$AdS_3$, $AdS_4$ and $AdS_5$ supersymmetric spacetimes in
eleven-dimensional supergravity using $G$-structures. These are obtained as
special cases of a class of supersymmetric $\bbR^{1,1}$, $\bbR^{1,2}$
and $\bbR^{1,3}$ geometries, naturally associated to 
M5-branes wrapping calibrated cycles in manifolds with $G_2$, $\SU(3)$ or $\SU(2)$ holonomy.
Specifically, the latter class is defined by requiring that
the Killing spinors satisfy the same set of projection conditions as
for wrapped probe branes, and that there is no electric flux. We show
how the R-symmetries of the dual field theories 
appear as isometries of the general $AdS$ geometries. We also show how known solutions
previously constructed in
gauged supergravity satisfy our more general $G$-structure conditions,
demonstrate that our conditions for half-BPS $AdS_5$ geometries are
precisely those of Lin, Lunin and Maldacena, and construct some
new singular solutions.


\end{quote}

\vfill

\end{titlepage}
\setcounter{equation}{0}


\section{Introduction}

M-theory on a supersymmetric background that contains an $AdS_{d+2}$
factor is expected to be dual to a $d+1$-dimensional superconformal
field theory (SCFT)~\cite{mald}. A key issue is thus both to
characterise the geometry of the generic eleven-dimensional
supergravity backgrounds of this type and to find explicit new
examples. In this paper we present a relatively simple way to
describe a large class of such spacetimes in terms of $G$-structures,
and show how some known solutions fit into this framework. We will
also present some new but singular solutions.

A number of results characterising $AdS$ solutions in M-theory have
already appeared in the literature. The generic
minimally supersymmetric backgrounds with an $AdS_3$ factor was analysed
in~\cite{J+D} (see also \cite{tsimpis}). 
Various authors have considered minimal $AdS_4$ compactifications; 
a general analysis is carried out in~\cite{L+S} 
and this is extended in~\cite{cvetic} 
(although our results differ slightly from
those in \cite{L+S}).
The generic $AdS_5$ case, dual to $\mathcal{N}=1$
SCFTs, was analysed  in~\cite{Gauntlett:2004zh}, while the $AdS_5$ case
dual to $\mathcal{N}=2$ SCFTs was analysed in~\cite{LLM}.

In this paper we will focus on $AdS$ solutions with no electric flux.
While this eliminates, for example, $AdS_4$ solutions of Freund-Rubin type, it
still includes rich classes of known solutions. One class was
originally derived from a gauged supergravity analysis. This work
began with the two $AdS_5$ solutions of~\cite{M-N} dual to
$\mathcal{N}=1$ and $\mathcal{N}=2$ SCFTs, and corresponding to
M5-branes wrapping holomorphic two-cycles in Calabi-Yau three-folds
and two-folds, respectively. This construction was extended to $AdS$
solutions corresponding to M5-branes wrapping various supersymmetric
cycles in~\cite{acharya,gkw,gk2} 
and including additional $AdS_3$ and $AdS_4$
examples with vanishing electric flux (for a review see~\cite{gcal}). A
second class arose from the general $AdS_5$ analysis
of~\cite{Gauntlett:2004zh}. By assuming that the compact internal
six-manifold is complex, an infinite number of new explicit solutions
were found. The six-manifolds were all $S^2$ bundles over a
K\"ahler-Einstein four manifold with positive curvature, or over a
product $S^2\times S^2$, $S^2\times T^2$, $S^2\times H^2$.  One
specific example in the $S^2\times H^2$ class gives the ${\cal N}=1$
$AdS_5$ solution of~\cite{M-N}.

It is natural, therefore, to (i)~characterise $AdS$ geometries
preserving various amounts of supersymmetry, (ii)~recover the known
wrapped brane solutions, (iii)~attempt to generalise them to new,
perhaps infinite, classes of solutions. In this paper we have achieved
(i) and (ii), and have initiated the analysis of (iii), by reducing
the general $G$-structure conditions to systems of first order ODEs, for
generalisations of known gauged supergravity solutions, and also for
some other special cases. While we have not found any new regular solutions,
physically one would expect many new
solutions, dual to SCFTs arising from wrapped branes, to be found in
the classes that we analyse, but we are agnostic as to whether or not
they can be found in explicit form.

Of course, from a more general perspective,  having a general
characterisation of the $AdS$ geometries is the first step in
developing existence theorems, which is a longer term goal. In
addition, the supersymmetric $AdS$ geometries we characterise here
will inherit rich structures from the dual SCFT, analogous to those
elucidated for the type IIB $D=5$ Sasaki-Einstein case in the
beautiful work of~\cite{msy,msy2}. It is also worth noting
that after analytic continuation the geometries we study should also
describe supersymmetric excitations of $AdS_{d+2}\times S^{9-d}$
geometries, analogous to the ``bubbling spacetimes'' discussed
in~\cite{LLM}.

Our approach is motivated by the general analysis of the
supersymmetric $AdS_5$ geometries in~\cite{Gauntlett:2004zh}. This was
carried out by directly studying the canonical $G$-structure specified
by the Killing spinors. However, it was also shown that the general
result could be obtained by an alternative strategy and this is
the one that we shall employ here. Using Poincar\'e coordinates, one
views the $AdS_{d+2}$ solution as a special case of an $\bbR^{1,d}$
solution. If one has a characterisation of the most general
supersymmetric $\bbR^{1,d}$ solution, one can therefore extract the
conditions for the most general $AdS_{d+2}$ solution as a special
case. In fact, to obtain the most general $AdS$ solution, it turns out
that it is not always necessary to start with the
most general supersymmetric $\bbR^{1,d}$ solution.

For the case of $AdS_5$ it is sufficient to start with supersymmetric
$\bbR^{1,3}$ geometries in a special class that we will call ``wrapped
brane geometries''. These geometries are characterised by the fact
that the Killing spinors are proportional to those preserved by a
probe M5-brane wrapping a supersymmetric cycle in a special holonomy
manifold: in particular they satisfy the same algebraic
conditions\footnote{These algebraic conditions can be equivalently
  phrased in terms of intersecting brane configurations.}. We
emphasise that the wrapped brane geometries need not contain
branes. We shall explain this in more detail in
section~\ref{wbs}. Following this we will present the necessary and
sufficient conditions for various wrapped brane geometries with
$\bbR^{1,1}$, $\bbR^{1,2}$ and $\bbR^{1,3}$ factors preserving various
amounts of supersymmetry. Using these results we are then able
to characterise the supersymmetric $AdS$ geometries.

At this point it is perhaps helpful to draw an analogy with
supersymmetric type IIB $AdS_5\times X_5$ solutions where $X_5$ is
Sasaki--Einstein. This class of solutions can be derived from the more
general class of supersymmetric solutions describing D3-branes that
are transverse to a Calabi--Yau three-fold. Specifically, one obtains
$AdS_5\times X_5$ in the special case that the Calabi--Yau is a
singular cone. This perspective plays a crucial role in identifying
the dual SCFTs which live on the D3-branes. The wrapped M5 brane
$\bbR^{1,d}$ solutions that we analyse here are the analogue of the
general IIB solutions describing D3-branes transverse to the
Calabi-Yau three-fold. Taking the special case which gives an $AdS_{d+2}$
factor is the analogue of requiring that the Calabi--Yau three-fold be a
cone. We similarly expect that, ultimately, this perspective will be
key in understanding the dual SCFTs as the decoupling limit of some
brane configuration. It is also worth noting that our approach can
equally be used to analyse supersymmetric $AdS$ solutions of any
supergravity theory.

A natural question to ask is whether the generic supersymmetric $AdS$
backgrounds can all be reproduced from the corresponding wrapped-brane
geometries. As we mentioned above this is certainly true for the
minimally supersymmetric $AdS_5$ case~\cite{Gauntlett:2004zh}. We will
see that it is also true for minimally supersymmetric $AdS_4$
spacetimes with vanishing electric flux, and very likely to be true
for $AdS_5$ spacetimes with additional supersymmetry. This at least
suggests that the 
wrapped-brane subclass is sufficiently general to give the generic
$AdS$ backgrounds with vanishing electric flux for all the cases we
consider. We will return to this question in our
conclusions.

The plan of the rest of this paper is as follows. In section~2, we
give a general discussion of the wrapped brane configurations that we
consider, together with an overview of their $G$-structures.
In section~3, we give the necessary and sufficient conditions for
supersymmetry of these
classes of wrapped brane solutions. These conditions can be neatly
understood using the generalised calibration conditions of~\cite{gp}
and this is discussed in section~4.

In section~5, we describe how we take the $AdS$ limit of the
wrapped brane metrics and in sections~6--8 we consider various
cases. Section~6 analyses $AdS_3$ solutions that are dual to
two-dimensional SCFTs with ${\cal N}=(2,0)$ and ${\cal N}=(4,0)$ supersymmetry, section~7
analyses  $AdS_4$ solutions that are dual to three-dimensional SCFTs
with ${\cal N}=1$ and $2$ supersymmetry, and section~8 analyses $AdS_5$
solutions that are dual to four-dimensional SCFTs with ${\cal N}=2$
supersymmetry (the ones with ${\cal N}=1$ supersymmetry are analysed
in~\cite{Gauntlett:2004zh}).

Section~9 discusses explicit examples of solutions of
the $AdS$ supersymmetry conditions. We provide a general discussion of
the $G$-structures underlying gauged supergravity $AdS$ solutions, and
show explicitly how they are realised for known examples. We highlight
potential generalisations of the gauged supergravity solutions,
discuss several other types of solutions, and explicitly construct new singular
$AdS_3$ solutions arising from branes wrapping K\"{a}hler four-cycles
in Calabi-Yau three-folds.

Section~10 concludes. We have relegated some technical material from
the main text to several appendices. In appendix~A we have listed the
spinor projections used to define the wrapped M5-brane geometries and
also the corresponding $G$-structures. Appendix~B gives more of the
general technical details involved in taking the $AdS$ limit
of the wrapped brane configurations. In appendix~C, we give a
representative example of the derivation of the $AdS$
supersymmetry conditions from the wrapped brane supersymmetry
conditions. In appendix~D, we prove that the supersymmetry conditions
for an M5 wrapping a K\"{a}hler two-cycle in a Calabi--Yau two-fold
descend, in the $AdS_5$ limit, precisely to the half-BPS
$AdS_5$ conditions of~\cite{LLM}. We will use the conventions
of~\cite{ggp} throughout this paper.

\section{Wrapped-brane spacetimes}
\label{wbs}

The main objective of the paper is to characterise supersymmetric
$AdS_{d+2}$ geometries that come from wrapped M5-brane geometries. The
wrapped brane geometries contain $\bbR^{1,d}$ factors and have the key
feature that they have Killing spinors that are proportional to those
preserved by a probe M5-brane wrapping a  supersymmetric cycle in a
special holonomy manifold (or equivalently by certain configurations
of intersecting M5-branes). In this section we will define this more
precisely.

Our ansatz for the wrapped M5-brane geometries starts with the general
ansatz for the metric given by
\begin{equation}
\label{g-ansatz}
   \dd s^2 = L^{-1}\, \dd s^2(\bbR^{1,d})
        + \dd s^2(\mathcal{M}_{10-d}) ,
\end{equation}
where the warp factor $L$ is a function of the coordinates on
$\mathcal{M}_{10-d}$. In terms of the wrapped probe-M5-brane
picture, the $\bbR^{1,d}$ factor corresponds to the unwrapped
world-volume of the M5-brane and $\mathcal{M}_{10-d}$ corresponds to
the wrapped M5-brane and the original geometry after the back-reaction has
switched on. Since M5-branes are sources of magnetic flux, we also
assume that the four-form flux $F$ lies solely in
$\mathcal{M}_{10-d}$, so that no electric flux, corresponding to
membranes, is present\footnote{For the cases of wrapped M5-branes that
  we will consider in this paper the electric flux is indeed vanishing
  for the explicitly known solutions of \cite{M-N,acharya,gkw,gk2}. However,
  there are additional solutions in \cite{gkw,gk2}, specifically for
  M5-branes wrapping 5-cycles or 4-cycles other than those considered
  here, where the M5-branes source electric flux and this would have
  to be properly taken into account in extending the analysis of this
  paper further.}.

The ansatz for the Killing spinors is best described by considering an
example. Let us take the case of a probe M5-brane wrapping a
co-associative cycle in $\bbR^{1,3}\times\mathcal{M}_{G_2}$, where
$\mathcal{M}_{G_2}$ is a seven-dimensional $G_2$-holonomy manifold. In
this case the unwrapped worldvolume is two-dimensional, so the metric
ansatz~\eqref{g-ansatz} has $d=1$. The two directions in $\bbR^{1,3}$
orthogonal to the world volume we refer to as ``overall transverse''
directions. Let us introduce a frame
\begin{equation}
   \dd s^2 = 2e^+e^- + (e^1)^2 + \dots + (e^9)^2,
\end{equation}
where $e^+$ and $e^-$ span the $\bbR^{1,1}$ unwrapped worldvolume
directions and $e^8$ and $e^9$ are the overall transverse directions.
Now consider the set of Killing spinors $\epsilon^i$. In
the case of the probe brane, the spacetime is
$\bbR^{1,3}\times\mathcal{M}_{G_2}$ and admits four Killing
spinors. The remaining seven basis one-forms can be chosen
such that these Killing spinors satisfy the eleven-dimensional
gamma-matrix projections
\begin{equation}
\label{g2-proj}
   \Gamma^{1234}\epsilon^i = \Gamma^{3456}\epsilon^i
     = \Gamma^{1357}\epsilon^i = - \epsilon^i .
\end{equation}
In this basis, the associative three-form $\Phi$ and co-associative
four-form $\Upsilon$ on $\mathcal{M}_{G_2}$, take the standard
form~\eqref{PhiUp}. If we now include the probe brane, the preserved
supersymmetries will be eigenspinors of the chirality operator on
the brane worldvolume. We can choose the orientation of the brane such
that this condition reads
\begin{equation}
\label{co-ass-proj}
   \Gamma^{+-1234} \epsilon^i = - \epsilon^i ,
\end{equation}
which is equivalent to $\Gamma^{+-}\epsilon^i=\epsilon^i$. This
reduces the number of Killing spinors to two. These can be
distinguished by, for instance, their eigenvalues under $\Gamma^9$
\begin{equation}
\label{eigenspinors}
   \Gamma^9\epsilon^1 = \epsilon^1, \qquad
   \Gamma^9\epsilon^2 = - \epsilon^2.
\end{equation}
%
Together the spinors define a $G_2$ structure on the
nine-dimensional manifold $\mathcal{M}_9=\mathcal{M}_{G_2}\times\bbR^2$.

The wrapped-brane spinor ansatz for the co-associative case is then
that we consider those spacetimes which admit a pair of Killing
spinors satisfying precisely the same projections as the probe brane geometry,
namely~\eqref{g2-proj},~\eqref{co-ass-proj}
and~\eqref{eigenspinors}. It is important to note that this is not the
most general ansatz for supersymmetric backgrounds with a warped
$\bbR^{1,1}$ factor and two Killing spinors (which are chiral in
$\bbR^{1,1}$, that is with $\mathcal{N}=(2,0)$ supersymmetry). The
calibration projections~\eqref{g2-proj} and~\eqref{co-ass-proj} could
be relaxed\footnote{For this particular case, but not the others we will
  consider, one particular way they can be relaxed corresponds to
  another wrapped M5-brane geometry, namely an M5-brane wrapping a
  K\"ahler four-cycle in a Calabi-Yau four-fold. This case also has
  $(2,0)$ supersymmetry in $\bbR^{1,1}$ but is distinguished from the
  co-associative case by the fact, for example, that both Killing
  spinors have the same eigenvalue under $\Gamma^9$. We do not treat
  this case in this paper because it is a case that should include
  electric flux.}, or, even if these
hold, there is no reason, a priori, that both $\epsilon^i$ are
eigenspinors of $\Gamma^9$.

In the co-associative example, note that one could choose the
Killing spinors such that both were simultaneous eigenspinors of the
five projection operators
\begin{equation}
\label{proj-set}
   \left\{
      \Gamma^{1234}, \Gamma^{3456}, \Gamma^{5678},
      \Gamma^{1357}, \Gamma^{+-}
   \right\} .
\end{equation}
This is actually characteristic 
of all the wrapped-brane Killing spinors we consider here
and is one way of defining the class. The cases that we shall consider
in this paper are summarised in Table~\ref{green}. The specific
projections and conventions for the various cases are given in
appendix~\ref{projs}.
\begin{table}[!th]
\begin{center}
\setlength{\tabcolsep}{0.45em}
\begin{tabular}{|c|c|c|c|}
\hline
wrapped brane & manifold & world-volume &
   supersymmetry \\
\hline
co-associative & $G_2$ holonomy & $\bbR^{1,1}$ & ${\cal N}=(2,0)$\\
K\"ahler 4-cycle & $CY_3$ & $\bbR^{1,1}$ & ${\cal N}=(4,0)$\\
associative & $G_2$ holonomy & $\bbR^{1,2}$ & ${\cal N}=1$\\
SLAG  & $CY_3$ & $\bbR^{1,2}$ & ${\cal N}=2$\\
K\"ahler 2-cycle & $CY_3$ & $\bbR^{1,3}$ & ${\cal N}=1$\\
K\"ahler 2-cycle & $CY_2$& $\bbR^{1,3}$ & ${\cal N}=2$\\
\hline
\end{tabular}
\end{center}
\caption{Wrapped M5-brane geometries and their supersymmetry}
\label{green}
\end{table}

At this point, we will summarise our ansatz for the class of wrapped
M5 brane spacetimes we consider. We demand that the metric contains a
warped $\mathbb{R}^{1,d}$ factor, and is of the form
(\ref{g-ansatz}). We demand that the flux has no electric components,
and so lies entirely in $\mathcal{M}_{10-d}$. Finally, we demand that
in each case the Killing spinors satisfy the appropriate probe brane
projections, and in particular, that they are simultaneous eigenspinors
of the five projection operators (\ref{proj-set}).
It is worth emphasising that because we are imposing these
projections, the $G$-structures that we use to characterise the
wrapped brane solutions below are in fact globally defined.

In what follows in the next two sections, the main point is to note
that there is a hierarchy of structures: structures with more supersymmetry
can be viewed in a simple way as pairs of structures with less
supersymmetry. In section~3, this will be used to derive the
conditions for supersymmetry for all these cases in a simple way,
starting only from the conditions for wrapping co-associative and
associative cycles. In section~4, we will discuss how the conditions
can be understood in terms of generalised calibrations. In section~5
we will start discussing how we derive the $AdS$ conditions.


\section{Supersymmetry conditions for wrapped M5-brane spacetimes}
\label{sec:susy}

In this section we will derive necessary and sufficient conditions on
$G$-structures for all the wrapped-brane geometries given in
Table~\ref{green}.
In fact, many of these conditions have been written down before
(although in some cases using stronger assumptions than ours). We
will provide a unified treatment, emphasizing the
feature that the structures with more supersymmetry can be viewed as
pairs of structures with less supersymmetry. Indeed we will see that
they all can be obtained from the associative and co-associative
cases. This is analogous to the fact that the conditions required on
$G$-structures for special holonomy manifolds in dimensions less than or
equal to eight can be obtained from multiple $Spin(7)$
structures. In the following section, we will also show how they can
be simply understood in terms of generalised calibrations. We will
organise the discussion by the dimension $d+1$ of the unwrapped brane
worldvolume.

\subsection{Co-associative cycles in $G_2$ holonomy
  ($\bbR^{1,1}$, ${\cal N}=(2,0)$)}
\label{co}

The general analysis of the conditions for supersymmetry for a pair or
spinors satisfying the projections~\eqref{g2-proj}
and~\eqref{co-ass-proj} was given in~\cite{oap}. Using these
conditions one can show that, assuming in addition only a warped
$\bbR^{1,1}$ factor, supersymmetry then implies that the Killing
spinors are eigenspinors of $\Gamma^{5678}$ and $F$ lies only in
$\mathcal{M}_9$ (i.e. there is no electric flux). In other words,
these parts of our ansatz need not be independently imposed for this
case. The Killing spinors define a preferred 
$(G_2\ltimes\mathbb{R}^7)\times\mathbb{R}^2$ structure in eleven
dimensions~\cite{nullstructure}. That is to say, at generic points
this is the stabilizer group of the two spinors.

The results of~\cite{oap} imply that the metric on $\mathcal{M}_9$ is compatible
with an integrable product structure (though the manifold is not necessarily a product)
\begin{equation}
\label{co-ass-g}
   \dd s^2(\mathcal{M}_9) = \dd s^2(\mathcal{M}_{G_2})
       + L^2(\dd y_1^2+\dd y_2^2),
\end{equation}
where both $L$ and $\Phi$ depend on all coordinates of $\mathcal{M}_9$ and
there is a $G_2$ structure $\Phi$ on $\mathcal{M}_{G_2}$. 
If we define
the orientation
\begin{equation}
\label{co-orient}
   \epsilon = \tfrac{1}{7} \Phi\wedge\Upsilon\wedge\vol_Y
\end{equation}
where $\vol_Y= L^2(\dd y_1\wedge\dd y_2)$, the remaining conditions for
supersymmetry may then be rewritten as
\bea\label{co1}
   {\vol_Y} \wedge \dd\Phi &=& 0,\\
   \dd(L^{-1}\Phi\wedge\Upsilon) &=&0,\\\label{co3}
   \Phi\wedge \dd\Phi&=&0,\\\label{co4}
   \star_9F&=&L\dd(L^{-1}\Upsilon).
\eea

We see that the metric~\eqref{co-ass-g} is conformally flat on the
overall transverse directions $\dd y_1$ and $\dd y_2$ (corresponding to
$e^8$ and $e^9$ in the conventions of appendix~\ref{projs}). Note that
these conditions imply the following useful expression for the flux
\begin{equation}
\label{coass-flux}
   (\star_2v) \wedge \,\dd\Phi = v \wedge F
\end{equation}
for any one-form $v$ lying in the overall transverse space and where
$\star_2$ is the Hodge star on the transverse space defined using the
orientation $\vol_Y$. As we will see in the next section, the
supersymmetry conditions \eqref{co1}-\eqref{co4} can be understood in
terms of generalised calibrations. This perspective also provides a
simple way to obtain~\eqref{coass-flux}.

\subsection{K\"{a}hler-4 cycles in $\SU(3)$ holonomy
  ($\bbR^{1,1}$, ${\cal N}=(4,0)$)}
\label{k4}

For a probe M5-brane wrapping a 
K\"ahler four-cycle in an $\SU(3)$ holonomy manifold,
the Killing spinors define a preferred 
$(SU(3)\ltimes\mathbb{R}^6)\times\mathbb{R}^3$ structure in eleven
dimensions \cite{nullstructure}.
From appendix~\ref{projs}, we see that the Killing spinor
projections~\eqref{SU3proj} and~\eqref{4in6proj}
are equivalent to a pair of co-associative projections with the $G_2$ structures
given by~\eqref{g2pair}. Thus the corresponding supersymmetry
conditions can be derived from the conditions~\eqref{co1}--\eqref{co4}
for the pair of $G_2$ structures $\Phi_\pm$.

After some manipulations one finds that the additional overall transverse
direction is given by $e^7=L\dd y_3$ so that the metric takes the
product form
\begin{equation}
\label{SLAG-g}
   \dd s^2(\mathcal{M}_9) = \dd s^2(\mathcal{M}_{\SU(3)})
       + L^2(\dd y_1^2+\dd y_2^2+\dd y_3^2) ,
\end{equation}
where there is an $\SU(3)$ structure on $\mathcal{M}_{\SU(3)}$ and $L$ depends
on all coordinates of $\mathcal{M}_9$. Fixing
the orientation
\begin{equation}
   \epsilon = \tfrac{1}{6}J\wedge J\wedge J \wedge \vol_Y ,
\end{equation}
with $\vol_Y=L^3\dd y_1\wedge\dd y_2\wedge\dd y_3$, the remaining
supersymmetry conditions read
\bea
{\vol_Y} \wedge  \dd(LJ) &=& 0,\\
\dd\Omega & = &0, \\
\star_9 F &=& L\dd\left(L^{-1}\tfrac{1}{2}J\wedge J\right).
\eea
These are in agreement with the results of \cite{kastor,Husain:2003df}
which were obtained with some additional assumptions about the form of
the backgrounds. It is worth noting that the conditions
imply that we have a K\"ahler metric on ${\cal M}_{SU(3)}$. The
complex structure is independent of $y_i$, while the K\"ahler form
$LJ$ is a function of the $y_i$.
Again one can also derive a useful expression for the flux
\begin{equation}
   (\star_3v) \wedge L^{-1}\dd(LJ) = - v \wedge F
\end{equation}
for any one-form $v$ lying in the overall transverse space spanned by
$\dd y_i$, and where $\star_3$ is defined using $\vol_Y$.

\subsection{Associative cycles in $G_2$ holonomy ($\bbR^{1,2}$,
  ${\cal N}=1$)}
\label{ass}

The general local analysis of the minimal supersymmetry conditions with a
warped $\bbR^{1,2}$ factor are given in~\cite{J+D}
(a discussion of some global issues can be found in \cite{tsimpis}). The Killing
spinors can always be chosen to satisfy the $G_2$ structure
conditions~\eqref{g2proj}. As shown in~\cite{J+D}, requiring $F$ to
lie on $\mathcal{M}_8$, i.e. no electric flux, then implies in
addition the associative calibration projection~\eqref{ass-proj} and
hence that we have a wrapped-brane geometry. The Killing spinors
define a preferred $G_2$ structure in eleven dimensions.

Summarising the conditions of~\cite{J+D} for supersymmetry in this
case, the metric is again a product
\begin{equation}
   \dd s^2(\mathcal{M}_8) = \dd s^2(\mathcal{M}_{G_2}) + L^2\dd y^2
\end{equation}
where $L$ depends on all coordinates on $\mathcal{M}_8$.
Fixing the orientation
\begin{equation}
   \epsilon = \tfrac{1}{7} \Phi\wedge\Upsilon\wedge v ,
\end{equation}
where $v=L\dd y$ is the overall transverse direction, we have
\bea
v\wedge \dd(L^{-1}\Upsilon)&=&0,\\
\dd(L^{-5/2}\Phi\wedge\Upsilon)&=&0,\\
\Phi\wedge \dd\Phi&=&0,\\
\star_8F&=&-L^{3/2}\dd(L^{-3/2}\Phi).
\eea
These imply the expression for the flux
\begin{equation}\label{bvccc}
   L\dd(L^{-1}\Upsilon) = - v\wedge F .
\end{equation}

\subsection{SLAG cycles in $\SU(3)$ holonomy ($\bbR^{1,2}$, ${\cal N}=2$)}
\label{SLAG3}

The supersymmetry conditions for this case are given
in~\cite{J+D}. Here we observe that they can be also obtained from our
co-associative conditions.

The spinor projections for a probe brane wrapping a SLAG cycle in an
$\SU(3)$-holonomy manifold~\eqref{SU3proj} and~\eqref{slagproj}
define a preferred $SU(3)$ structure in eleven dimensions.
They are equivalent to a pair of co-associative
projections. From~\eqref{slagproj}, one sees the two corresponding
co-associative structures are given by $G_2$-structures
$\Phi_\pm$~\eqref{g2pair}, together with exchanging $e^+$ and $e^-$.

Since the one-form $e^7$ lies along the unwrapped worldvolume, we
demand that $e^7=L^{-1/2}\dd x_2$ and in addition that the flux has no
component in this direction. Then demanding that the two $G_2$
structures satisfy the supersymmetry conditions
\eqref{co1}--\eqref{co4}, we recover the results of~\cite{J+D}. In
particular, we find that the eight-dimensional metric is a product
\begin{equation}
   \dd s^2(\mathcal{M}_8) = \dd s^2(\mathcal{M}_{\SU(3)})
      + L^2 (\dd y_1^2 + \dd y_2^2 ) 
\end{equation}
where $L$ depends on all coordinates on $\mathcal{M}_8$.
Defining the orientation
\begin{equation}
   \epsilon = \tfrac{1}{6}J \wedge J \wedge J \wedge \vol_Y ,
\end{equation}
where $\vol_Y=L^2\dd y_1\wedge\dd y_2$, the remaining supersymmetry
conditions read
\bea\label{219}
  {\vol_Y}\wedge \dd\im\Omega &=& 0,\\\label{221}
  \dd(L^{-1/2}J) &=& 0,\\\label{220}
  \dd\re\Omega\wedge\re\Omega &=& 0,\\\label{222}
   \star_8F &=& L^{3/2}\dd(L^{-3/2}\re\Omega).
\eea
These imply the expression for the flux
\begin{equation}
\label{rrrr}
   (\star_2v) \wedge \dd\im\Omega = v \wedge F ,
\end{equation}
for any $v$ lying in the overall transverse space.

\subsection{K\"{a}hler-2 cycles in $\SU(3)$ holonomy ($\bbR^{1,3}$,
  ${\cal N}=1$)}

For the case of an M5-brane wrapping a K\"{a}hler 2-cycle in an
$\SU(3)$-holonomy manifold, the Killing spinors define a preferred
$SU(3)$ structure in eleven dimensions. As in the SLAG case, we
may derive the supersymmetry conditions directly from the conditions
corresponding to wrapping associative cycles. The two-cycle spinor
projections~\eqref{SU3proj} and~\eqref{2in6proj} are equivalent to a
pair of associative projections with $G_2$ structures given
by~\eqref{g2pair}. Since $e^7$ lies along the unwrapped
worldvolume, we demand that $e^7=L^{-1/2}\dd x_3$ and we also demand that
the flux has no component along $e^7$.

One finds the product metric
\begin{equation}
   \dd s^2(\mathcal{M}_7) = \dd s^2(\mathcal{M}_{\SU(3)})
      + L^2 \dd y^2 
\end{equation}
where $L$ depends on all coordinates on $\mathcal{M}_7$.
Defining the orientation
\begin{equation}
   \epsilon = \tfrac{1}{6}J \wedge J \wedge J \wedge v ,
\end{equation}
where $v=L\dd y$, the remaining supersymmetry conditions read
\bea
   v\wedge \dd(L^{-1}J\wedge J) &=& 0,\\
   \dd(L^{-3/2}\Omega) &=& 0,\\
   \star_7F &=& -L^2\dd(L^{-2}J).
\eea
These are consistent with the results of~\cite{smith1}, where
additional assumptions about the form of the background were made. Note that
these conditions imply that  ${\cal M}_{SU(3)}$ is a
complex manifold, with a complex structure independent of $y_i$ and
an hermitian metric dependent on $y_i$. The conditions also
imply
\begin{equation}
   \tfrac{1}{2}L\dd\left(L^{-1}J\wedge J\right) = - v \wedge F .
\end{equation}

\subsection{K\"{a}hler-2 cycles in $\SU(2)$ holonomy ($\bbR^{1,3}$,
  ${\cal N}=2$)}

The final case we consider is that of an M5-brane wrapping
a K\"ahler two-cycle in an $\SU(2)$-holonomy manifold.
The Killing spinors for this case define a preferred 
$SU(2)$ structure in eleven dimensions.
The spinor projections ~\eqref{SU2proj} and~\eqref{2in4proj} are
equivalent to a pair of K\"ahler two-cycle in $\SU(3)$ holonomy
projections, with the two corresponding $\SU(3)$ structures given
by~\eqref{SU3pair}.

Using the conditions for supersymmetry for each of the $\SU(3)$
wrapped brane geometries that were derived in the last subsection,
one can show that $e^5=L\dd y_2$ and
$e^6=L\dd y_3$ so that the seven-dimensional metric is a product
\begin{equation}
   \dd s^2(\mathcal{M}_7) = \dd s^2(\mathcal{M}_{\SU(2)})
      + L^2 (\dd y_1^2 + \dd y_2^2 + \dd y_3^2) 
\end{equation}
where $L$ depends on all coordinates on $\mathcal{M}_7$.
Defining the orientation
\begin{equation}
   \epsilon = \tfrac{1}{2}J^1 \wedge J^1 \wedge \vol_Y ,
\end{equation}
where $\vol_Y=L^3\dd y_1\wedge\dd y_2\wedge\dd y_3$, the remaining
supersymmetry conditions read
\bea
  \dd(L^{-1/2}J^2) = \dd(L^{-1/2}J^3) &=& 0,\\
  {\vol_Y}\wedge \dd(LJ^1) &=& 0,\\
   \star_7F& = &L^2\dd(L^{-2}J^1) .
\eea
These conditions were first derived in~\cite{fayy3} (extending the results of
\cite{smith}). The combination $J^3+\ii J^2$ defines a complex
structure on ${\cal M}_{SU(2)}$ independent of $y_i$, while $LJ^1$
defines a K\"ahler metric at fixed $y_i$. The conditions also imply
\begin{equation}
   (\star_3 v)\wedge L^{-1}\dd(L J^1) = - v\wedge F ,
\end{equation}
where $v$ is any one-form in the overall transverse space.


\section{Relation to generalised calibrations}
\label{sec:gen-cal}

A key early paper highlighting the role of generalised
calibrations~\cite{Gutowski:1999tu} in describing classes of
supersymmetric supergravity geometries is the work by Cho et
al.~\cite{kastor}. This was subsequently explored in the context of
IIB supergravity in~\cite{Gauntlett:2001ur}. In~\cite{ggp} it was
shown that generic eleven-dimensional supersymmetric solutions admit
generalised calibrations. The fact that, for certain cases of
solutions, {\it all} of the conditions for supersymmetry can be
understood in terms of generalised calibrations was discussed
in~\cite{int-tor}. For some of the cases that we consider in this
paper, the observation that a subset of the supersymmetry conditions
are related to generalised calibrations was made
in~\cite{Husain:2003df,fayy3,Fayyazuddin:2005ds}. The relation to
generalised calibrations of warped supersymmetric geometries of the
form $\bbR^{1,2}\times\mathcal{M}_8$ was discussed in detail
in~\cite{J+D}. In this section we will briefly show that in fact all
of the supersymmetry conditions for the wrapped brane geometries can
be interpreted this way.

A calibrating $p$-form $\Xi$ on a Riemannian manifold $\mathcal{M}$
has the property that the metric-induced volume form on any oriented
$p$-dimensional subspace $\xi$ of $T_x\mathcal{M}$ is greater than or
equal to the restriction of $\Xi$ to $\xi$. A
$p$-dimensional submanifold $C_p$ is calibrated if the bound is
saturated, $\Xi|_{T_xC_p}=\vol_{T_xC_p}$, everywhere on
$C_p$. Conventionally the calibrating form is also required to be
closed. This then implies that the calibrated cycle has minimum volume
in its homology class. Crucially all the structure forms defining
special holonomy manifolds are calibrating forms. The different
supersymmetric cycles (associative, co-associative, K\"ahler, SLAG) we
have been discussing are all calibrated cycles.

For a generalised calibration~\cite{Gutowski:1999tu}, one retains the
algebraic condition relating $\Xi$ to the volume, but generalises the
differential condition so that $\Xi$ is no longer closed. Instead
$\dd\Xi$ involves the flux, and any warping factor if the spacetime
is a warped product $\bbR^{1,d}\times\mathcal{M}$ as
in~\eqref{g-ansatz}. The point is that the calibrated cycles now
extremize not their volume but rather the brane energy, including for
instance the contribution from the flux, for probe branes wrapping the
corresponding cycle. In the case of M5-branes, one gets conditions like
\be
\label{gen-cal}
   L^{m} \dd \left( L^{-m} \Xi \right) = \star_{10-d} F ,
\ee
for some $m\in\mathbb{Q}$. In a supersymmetric background, the
generalised calibrated cycles are supersymmetric.

Note that for each of the wrapped brane geometries, the supersymmetry
conditions contain one condition of the form~\eqref{gen-cal}. In fact,
the remaining supersymmetry conditions can also
be interpreted as generalised calibrations related to other ways in
which probe M5-branes (or M2-branes) can wrap various calibrated
cycles whilst preserving supersymmetry.  (In the context of type II
supergravity this is discussed in more detail in the introduction and
conclusion of~\cite{int-tor}). To see this we will show, equivalently,
that the supersymmetry conditions can be obtained from the generalised
calibration conditions arising from minimally supersymmetric solutions
of $D=11$ supergravity~\cite{ggp}.

The essential point is that the assumption that each Killing spinor is
an eigenspinor of the five projection operators~\eqref{proj-set}
implies that each spinor defines a local
$(\Spin(7)\ltimes\bbR^8)\times\bbR$ structure, with the structures
fitting together in a very simple way. Explicitly,
following~\cite{ggp}, if $\epsilon$ has eigenvalue $+1$ under
$\Gamma^{+-}$ and $-1$ under all the other projectors, the
corresponding structure can be written as
\begin{equation}
\label{spin7}
\begin{aligned}
   K &= e^+ , \\
   \Omega &= K \wedge v , \\
   \Sigma &= K \wedge \phi ,
\end{aligned}
\end{equation}
where $v=e^9$ and $\phi$ is the $\Spin(7)$ invariant
\begin{equation}
\begin{aligned}
   \phi &= {} - e^{1234} - e^{1256} - e^{1278} - e^{3456}
      - e^{3478} - e^{5678} \\
      & \qquad - e^{1357} + e^{1368} + e^{1458}
      + e^{1467} + e^{2358} + e^{2367} + e^{2457} - e^{2468}.
\end{aligned}
\end{equation}
One can then show~\cite{ggp}, that the Killing spinor equation for
$\epsilon$ implies set of differential conditions on
$(K,\Omega,\Sigma)$,
\begin{equation}
\label{cal-cond}
\begin{aligned}
   \dd K &= \tfrac{2}{3}i_\Omega F + \tfrac{1}{3}i_\Sigma \star F , \\
   \dd \Omega &= i_K F, \\
   \dd \Sigma &= i_K \star F - \Omega \wedge F .
\end{aligned}
\end{equation}
One can view these as a set of generalised calibration
conditions. Since $K$ is null, the first one is associated to massless
particles, the second is associated to wrapped M2-branes (coupling to
electric flux) and the third to wrapped M5-branes (coupling to
magnetic flux).

Now consider, for instance, the case of a co-associative
calibration. We now have a pair of Killing spinors. Following our
discussion in section~\ref{wbs}, each has positive eigenvalue under
$\Gamma^{+-}$ and negative eigenvalue under $\Gamma^{1234}$,
$\Gamma^{3456}$ and $\Gamma^{1357}$. They are distinguished by their
eigenvalue under $\Gamma^{5678}$ or equivalently $\Gamma^9$. From this
perspective, each spinor defines a different
$(\Spin(7)\ltimes\bbR^8)\times\bbR$ local structure. Explicitly, these
are given by~\eqref{spin7} with $\{K=e^+,\phi=\phi_\pm,v=\pm e^9\}$
where
\begin{equation}
\label{phi-pm}
   \phi_\pm = \mp \Phi \wedge e^8 - \Upsilon ,
\end{equation}
with $\Phi$ the three-form defining the $G_2$ structure. In fact, more
generally, by taking constant linear combinations of $\epsilon^1$ and
$\epsilon^2$ we get a family of $(\Spin(7)\ltimes\bbR^8)\times\bbR$
structures $\{K=e^+,\phi=\phi(\theta),v=v(\theta)\}$ where
\begin{equation}
\begin{aligned}
   v(\theta) &= \cos\theta e^9-\sin\theta e^8 , \\
   \phi(\theta) &= - \Phi \wedge (\cos\theta e^8+\sin\theta e^9)
      - \Upsilon .
\end{aligned}
\end{equation}
and $\theta$ is constant.

Supersymmetry implies that the generalised calibration
conditions~\eqref{cal-cond} must be satisfied for all these
structures. This then gives us a simple way to derive, in this case,
the wrapped-brane geometry supersymmetry conditions for a
co-associative calibration. Explicitly we start with the metric
ansatz~\eqref{g-ansatz}. Writing $K=e^+=L^{-1}\dd x^+$ and given that the
flux lies solely on $\mathcal{M}_9$, the M2-brane calibration
condition for $\Omega$ with general $\theta$ implies that
\begin{equation}
\label{M2-cal}
   \dd(L^{-1}v(\theta)) = 0
\end{equation}
or locally $e^8=L\dd y_1$ and $e^9=L\dd y_2$. Given the
orientation~\eqref{co-orient}, the M5-brane $\Sigma$ calibration
condition gives
\begin{equation}
\label{M5-cal}
\begin{aligned}
   L \dd(L^{-1}\Upsilon) &= \star_9 F , \\
   L \dd (L^{-1} v(\theta) \wedge \Phi)
      &= - \star_2v(\theta) \wedge F ,
\end{aligned}
\end{equation}
for all $\theta$. After some manipulations one can show that these
imply the set of supersymmetry conditions~\eqref{co1}--\eqref{co4}
given in the previous section. Note that in this case the $K$
calibration condition is implied by the M2-brane and M5-brane
calibration conditions.

Thus we see that the calibration conditions~\eqref{cal-cond} for each
spinor are in fact necessary and sufficient for supersymmetry of the wrapped
M5-brane geometry. A similar calculation can be used to derive the
supersymmetry conditions for the wrapped-brane geometries related to
associative calibrations. We saw in the previous section that all the
other wrapped-brane geometry supersymmetry conditions could be derived
from this basic pair, and hence, ultimately from the calibration
conditions~\eqref{cal-cond} (in fact limited only to the $\Omega$ and
$\Sigma$ calibrations). From this perspective, the conditions for
supersymmetry are equivalent to requiring that all the possible
structure forms, compatible with the $G$-structure of the background,
are actually generalised calibrations.

Physically one can view the set of calibration conditions as
corresponding to all the possible additional supersymmetric wrapped
probe M5-branes and probe M2-branes compatible with the supersymmetry
of the wrapped-brane geometry. In the example above the first
conditions~\eqref{M2-cal} correspond to calibration conditions for
M2-branes spanning $e^+$, $e^-$ and $v(\theta)$. The second set of
conditions~\eqref{M5-cal}, correspond to M5-branes spanning $e^+$,
$e^-$ and a co-associative cycle in $\mathcal{M}$, or $e^+$, $e^-$,
$v(\theta)$ and an associative cycle in $\mathcal{M}$.


\section{$AdS$ spacetimes from wrapped-brane spacetimes}
\label{sec:red}

In this section, we will discuss how to obtain
$AdS$ backgrounds from the wrapped-brane geometries discussed
thus far. Once we have formulated the
$AdS$ limit, we may simply insert it in the wrapped-brane
supersymmetry conditions to obtain the conditions for supersymmetry of
the $AdS$ spacetimes.

In Poincar\'e coordinates a general $AdS_{d+2}$ spacetime can be written as
\begin{equation}
\label{AdSmetric}
\begin{aligned}
   \dd s^2 &= \lambda^{-1} \dd s^2(AdS_{d+2})
         + \dd s^2(\mathcal{N}_{9-d}) \\
      &= \lambda^{-1} \left[ \ee^{-2mr} \dd s^2(\bbR^{1,d})
         + \dd r^2 \right] + \dd s^2(\mathcal{N}_{9-d}),
\end{aligned}
\end{equation}
This can be obtained from the wrapped geometries~\eqref{g-ansatz} by
demanding
\begin{equation}
\label{Ads-red}
\begin{aligned}
   L &= \ee^{2mr} \lambda , \\
   \dd s^2(\mathcal{M}_{10-d})
      &= \lambda^{-1} \dd r^2 + \dd s^2(\mathcal{N}_{9-d}) ,
\end{aligned}
\end{equation}
and where the warp factor $\lambda$ is taken to be a function of the
coordinates on $\dd s^2(\mathcal{N}_{9-d})$.
Note that the vector $\der/\der r$ is both Killing and
hypersurface orthogonal on $\dd s^2(\mathcal{M}_{10-d})$. We also
assume that the flux $F$ lies solely in $\mathcal{N}_{9-d}$, and is
independent of the $AdS$ radial coordinate, so that the full solution
preserves the $AdS$ isometries.

To analyse this reduction of $\mathcal{M}_{10-d}$ to
$\mathcal{N}_{9-d}$, we note that in all the wrapped-brane geometries
the metric took the particular product form where the overall
transverse directions are conformally flat,
\begin{equation}
\label{M-split}
   \dd s^2(\mathcal{M}_{10-d}) =
       \dd s^2(\mathcal{M}_G)
          + L^2 \left( \dd t^2 + t^2 \dd\Omega_{q-1}^2 \right) ,
\end{equation}
where we have introduced polar coordinates on the $q$-dimensional
transverse space, so $\dd\Omega^2_{q-1}$ is the round metric on the unit
sphere $S^{q-1}$. For the cases of interest $q=1,2,3$. Note in
addition, that there is $G$-structure on $\mathcal{M}_G$ in each
case. This product structure means that generically the radial
vector $\der/\der r$ will split into a part in $\mathcal{M}_G$ and a
part in the overall transverse space. In particular, we can write the
$AdS$ unit radial one-form as
\begin{equation}
\label{r-decomp}
   \lambda^{-1/2} \dd r = \sin\theta\,\hat{u} + \cos\theta\,\hat{v},
\end{equation}
where $\hat{u}$ is a unit one-form in $\mathcal{M}_G$, and
$\hat{v}$ is a unit one-form in the overall transverse space.

We will make the assumption that $\hat{v}$ is given by
\begin{equation}
\label{t-ansatz}
   \hat{v} = L\dd t,
\end{equation}
and so lies along the radial direction $t$ of the conformally flat
overall transverse space. In addition we will assume that the rotation
angle $\theta$ must be independent of the $AdS$ radial coordinate. As
we will see below, these assumptions lead to geometries with at least
part\footnote{For the cases we consider of M5-branes wrapping K\"ahler cycles,
the complex structure on $\mathcal{M}_G$ acting on $\hat u$ picks out another
direction in $\mathcal{M}_G$ and this also contributes to
the $R$-symmetry.}
of the $R$-symmetry of the field theory realised as
isometries of the sphere $S^{q-1}$, as one might expect. 
Here we have presented (\ref{t-ansatz})
and the $r$-independence of $\theta$ as assumptions, though we
emphasise that rather stronger statements regarding the generality, or
otherwise, of our $AdS$ limit can be made. As we discuss in more detail in
appendix~\ref{appA}, for the case of one overall transverse direction,
the rotation angle $\theta$ is in fact necessarily independent of
$r$, so in this case our $AdS$ limit is in fact the most general way
of obtaining an $AdS$ geometry from the wrapped brane spacetime.  For
the case of two or three overall 
transverse directions our results are slightly weaker, but we show
that with a suitable assumption of $r$-independence
of the frame rotation, the part of the $AdS$ radial direction which
lies in the overall transverse space must in fact lie entirely along the radial
direction of the overall transverse space, as in (\ref{t-ansatz}).

Now, introducing the orthogonal combination
\begin{equation}
   \hat{\rho} = \cos\theta\,\hat{u} - \sin\theta\,\hat{v} ,
\end{equation}
the fact that $\dd t$ is closed, and $\theta$ is independent of $r$,
then implies that
\begin{equation}
   \hat{\rho} =\frac{\lambda}{2m\sin\theta}\dd(\lambda^{-3/2}\cos\theta).
\end{equation}
Defining a new coordinate $\rho=\lambda^{-3/2}\cos\theta$ one then has
the relation $t=-(\rho/2m)\ee^{-2mr}$ and
\bea
   \hat{\rho}& = &\frac{\lambda \dd\rho}{2m\sqrt{1-\lambda^3\rho^2}},\nn
   \hat{u}&=&\l^{-1/2}\sss dr+\frac{\l^{5/2}\r d\r}{2m\sss}.
\eea
Extracting the $AdS$ factor, one finds that the metric $\dd
s^2(\mathcal{N}_{9-d})$ then takes the form
\begin{equation}
\label{N-metric}
   \dd s^2(\mathcal{N}_{9-d}) =
      \dd s^2(\mathcal{M}_{G'})
         + \frac{\lambda^2}{4m^2}\left(
            \frac{\dd\rho^2}{1-\lambda^3\rho^2}
            + \rho^2 \dd\Omega_{q-1}^2 \right) ,
\end{equation}
where $\dd s^2(\mathcal{M}_{G'})$ is defined via
\begin{equation}
   \dd s^2(\mathcal{M}_G) = \dd s^2(\mathcal{M}_{G'})
      + \hat{u}\otimes \hat{u} .
\end{equation}
The $G'$-structure on $\mathcal{M}_{G'}$ is a reduction of the
$G$-structure on $M_G$, defined by picking out the particular one-form
$\hat{u}$. 
It is useful in what follows to define a
normalised volume form on the transverse sphere $S^{q-1}$
in~\eqref{Ads-red}
\begin{equation}\label{normvol}
   \hvol(S^{q-1}) =
      \left(\frac{\lambda\rho}{2m}\right)^{q-1} \vol(S^{q-1}),
\end{equation}
where $\vol(S^{q-1})$ is the volume form on the unit sphere.

Given the supersymmetry conditions on the original space
$\mathcal{M}_{10-d}$ it is then straightforward to take~\eqref{Ads-red}
with $\dd s^2(\mathcal{N}_{9-d})$ given
by~\eqref{N-metric}, demand that the flux has no components along the
AdS radial direction, and hence derive the supersymmetry conditions for an
$AdS_{d+2}$ geometry in terms of the $G'$-structure. It is worth noting that this $G'$-structure is,
in general, only locally defined, since there can be points 
where $\sin\theta=0$ and hence the vector $\hat u$ is ill-defined.

The discussion thus far has been for the generic case where $\der/\der
r$ lies partly in $\mathcal{M}_G$ and partly in the overall transverse
space. There are two special cases one should also consider. First
$\der/\der r$ could lie entirely in $\mathcal{M}_G$. 
This is excluded since it is inconsistent with $\der/\der r$ being
Killing, since from~\eqref{Ads-red} and~\eqref{M-split} we see that the
overall transverse space would then have an explicit dependence on $r$.

The second possibility is that $\der/\der r$ lies entirely in the
overall transverse space. Note that in all the cases analysed in
section~\ref{sec:susy} we have a condition of the form
\begin{equation}
\label{vol-cond}
   \dd ( L^m \vol_G ) = 0 ,
\end{equation}
for some $m\in\mathbb{Q}$, where $\vol_G$ is the volume form on
$\mathcal{M}_G$. Since $\der/\der r$ is Killing and assuming it lies
solely in the overall transverse space, we have that $\vol_G$ is
independent of $r$. This implies, provided $m\neq 0$, that
\begin{equation}
   \dd r \wedge \vol_G = 0,
\end{equation}
which is impossible. Note that $m=0$ only in the case of K\"ahler-4
calibrations in $\SU(3)$ holonomy. Thus only in this one special
case do we need to consider the case where $\der/\der r$ lies solely
in the overall transverse space. This is discussed separately in
section~\ref{AdS-4in6}.

Let us end this section by noting that for all the $AdS$ geometries
we obtain from the wrapped-brane spacetimes, supersymmetry implies that
all equations of motion and the Bianchi identity are identically
satisfied. From section~\ref{sec:susy} it is clear that for all
wrapped-brane geometries we have
\begin{align}
   \star_{10-d} F &= L^{r_1} \dd \left( L^{-r_1} \Xi_1 \right) ,
      \label{*F} \\
   v \wedge F &=
      (\star_p v) \wedge L^{r_2} \dd \left( L^{-r_2} \Xi_2 \right) ,
      \label{F}
\end{align}
for some $r_1,r_2\in\mathbb{Q}$, some calibration forms $\Xi_1$ and
$\Xi_2$, and any one-form $v$ in the overall transverse space. By
taking the exterior derivative of~\eqref{*F}, one automatically
satisfies the equation of motion for $F$ for any wrapped-brane
geometry. Generically, the Bianchi identity is not satisfied as a
consequence of the wrapped brane supersymmetry conditions, and must be
imposed. Imposing the Bianchi identity, the results of \cite{gp},
\cite{oap}, then imply that the Einstein equations are identically
satisfied, with the possible exception of the $++$ component in the
co-associative and K\"{a}hler-4 cases; however because we have assumed
a warped Minkowski factor, it is easy to check that these components
are in fact satisfied. Thus to guarantee a solution of the field
equations for the wrapped brane spacetimes, we need only impose the
Bianchi identity in addition to the supersymmetry conditions. By
taking the $AdS$ limit of~\eqref{F}, we may easily deduce
the flux in each case, and in each case we have verified that the
Bianchi identity for the flux is identically satisfied in the $AdS$
limit. Therefore the supersymmetry conditions in the $AdS$ limit are
necessary and sufficient to guarantee a solution of all the field
equations and the Bianchi identity.


In the following sections, we present the supersymmetry conditions for
$AdS_{d+2}$ spacetimes, using the reduction~\eqref{Ads-red}, for each of
the different wrapped brane geometries. The derivations are straightforward
but a bit long, so we just give some sample calculations for a
representative example in appendix C.


\section{Supersymmetric $AdS_3$ spacetimes}

In this section, we will use the reduction discussed in the previous
section to obtain the conditions for the supersymmetric $AdS_3$
spacetimes contained in the wrapped brane geometries with an
$\mathbb{R}^{1,1}$ factor. Specifically, these will correspond to
M5-branes wrapping co-associative cycles in $G_2$-holonomy manifolds
and K\"{a}hler four-cycles in $\SU(3)$-holonomy manifolds.


\subsection{$AdS_3$ spacetimes from wrapping co-associative cycles}

These geometries will be dual to two-dimensional SCFTs with a chiral
$\mathcal{N}=(2,0)$ supersymmetry, and hence with a $U(1)$ R-symmetry.

The overall transverse space is two-dimensional in this case, and so
we have $q=2$ in~\eqref{M-split}. Picking the unit one-form $\hat{u}$ in
$\mathcal{M}_{G_2}$ breaks the local structure to $G'=\SU(3)$, defined
by $J$ and $\Omega$, as given in appendix~\ref{projs} with
$e^7=\hat{u}$. Thus we have
\begin{equation}
   \dd s^2(\mathcal{N}_8) =
      \dd s^2(\mathcal{M}_{\SU(3)})
         + \frac{\lambda^2}{4m^2}\left(
            \frac{\dd\rho^2}{1-\lambda^3\rho^2}
            + \rho^2 \dd\phi^2 \right) ,
\end{equation}
where $\phi$ is a coordinate on the $S^1$. We define the orientation
\begin{equation}
   \epsilon = \tfrac{1}{6} J\wedge J\wedge J
       \wedge \hat{\rho} \wedge \hvol(S^1) ,
\end{equation}
where we recall that the normalised volume $\hvol(S^1)$ was defined in \eqref{normvol}.
The remaining independent conditions for supersymmetry turn out to be as follows:
\begin{align}
   \dd\left(\frac{1}{\lambda^{3/2}\rho}J\wedge\hat{\rho}
      - \im\Omega \right) &=0 , \label{44} \\
   \dd\left(\frac{1}{2\l}J\wedge J
      + \l^{1/2}\r\re\Omega\wedge\hat{\rho}\right) &=0 . \label{45}
\end{align}
From the flux condition~\eqref{coass-flux}, we find
\begin{equation}
   F = - \frac{1}{\l\r}\hvol(S^1)\wedge
      \dd\left(\l^{-1/2}\sqrt{1-\l^3\r^2}J\right) .
\end{equation}
It follows from the supersymmetry conditions~\eqref{44} and~\eqref{45}
that the rotational Killing vector on $S^1$ is a Killing vector of the full
solution that preserves the flux. 
Therefore the supersymmetry conditions imply that
generically the isometry group of the $AdS$ limit, and hence the
R-symmetry of the general dual SCFT, is $U(1)$, as expected. In
section~\ref{solutions} we will recover the explicit supersymmetric
solution of~\cite{gkw} using these results.

\subsection{$AdS_3$ spacetimes from wrapping K\"{a}hler four-cycles in
  $\SU(3)$ holonomy}
\label{AdS-4in6}

For the case of wrapped brane geometries corresponding to K\"{a}hler
four-cycles in spaces with $\SU(3)$ holonomy there are three overall
transverse directions and so $q=3$. As we discussed in the previous
section, there are two distinct ways of taking the $AdS_3$ limit
and we shall discuss both of them. The $AdS_3$ geometries will be dual
to two-dimensional SCFTs with a chiral $\mathcal{N}=(4,0)$
supersymmetry.

\paragraph{$AdS$ radial direction from the overall transverse space}
Demanding that the $AdS$ radial direction lies entirely in the overall
transverse space implies that $\lambda$ is constant and that
$\der/\der r$ lies along the radial direction of the overall
transverse space. Hence, rescaling so $\lambda=1$, instead
of~\eqref{N-metric}, we have
\begin{equation}
\label{CYsol}
   \dd s^2(\mathcal{N}_8) = \dd s^2(\mathcal{M}_{\SU(3)})
       + \frac{1}{4m^2} \dd \Omega^2_2 .
\end{equation}
The supersymmetry conditions then imply that
\begin{equation}
   \dd J = \dd\Omega = 0 ,
\end{equation}
and $F=-\vol(S^2)\wedge J/2m$. Therefore $\mathcal{M}_{\SU(3)}$ is
Calabi--Yau and, in addition, the rotational Killing vectors on the $S^2$
are Killing vectors of the full solution and also preserve the flux. This is the well
known $AdS_3\times S^2\times CY_3$ solution. Generically, the
isometry group of the space transverse to the $AdS$ factor is
$\SU(2)$.

\paragraph{Generic $AdS$ radial direction}
Generically the $AdS$ radial direction has a component in
$\mathcal{M}_{\SU(3)}$ and a component in the overall transverse
space as discussed in section~\ref{sec:red}. The component $\hat{u}$
in $\mathcal{M}_{\SU(3)}$ reduces the structure to $G'=\SU(2)$ in
five dimensions, where in the conventions of appendix~\ref{projs} we
have $e^6=\hat{u}$. Such a structure is defined by a triplet of
two-forms $J^i$, defining a conventional four-dimensional $\SU(2)$
structure together with an additional one-form $\hat{w}=e^5$ (in the
conventions of appendix~\ref{projs}). Thus we have
\begin{equation}
   \dd s^2(\mathcal{N}_8) = \dd s^2(\mathcal{M}_{\SU(2)})
      + \hat{w}\otimes\hat{w}
      + \frac{\lambda^2}{4m^2}\left(
            \frac{\dd\rho^2}{1-\lambda^3\rho^2}
            + \rho^2 \dd\Omega_2^2 \right) ,
\end{equation}
where there is an $\SU(2)$ structure on the four-dimensional space
$\mathcal{M}_{\SU(2)}$.

Defining the orientation
\bea
\epsilon = \tfrac{1}{6}J^i\wedge J^i \wedge \hat{w}
    \wedge \hat{\rho} \wedge \hvol(S^2),
\eea
the conditions for supersymmetry are
\bea\label{411}
\dd\left(\l^{-1/2}\sss J^2\right) &=&0,\\\label{412}
\dd\left(\l^{-1/2}\sss J^3\right)&=&0,\\\label{413}
\dd\left(\l\r J^1+\l^{-1/2}\hat{w}\wedge\hat{\rho}\right)
   &=&0,\\\label{414}
J^3\wedge\dd\left(\frac{\l^{1/2}}{\sss}\hat{w}\right)
   &=&J^2\wedge\dd\left(\frac{\l^2\r}{\sss}\hat{\rho}\right)
   ,\\\label{415}
J^2\wedge\dd\left(\frac{\l^{1/2}}{\sss}\hat{w}\right)
   &=&-J^3\wedge\dd\left(\frac{\l^2\r}{\sss}\hat{\rho}\right) ,
\eea
and the flux is given by
\bea
F=-\frac{1}{\l^2\r^2}\hvol(S^2)\wedge\left[
   \dd\left(\l^{1/2}\r\sss\hat{w}\right)
   +2m\left(\l\r J^1+\l^{-1/2}\hat{w}\wedge\hat{\rho}\right)
   \right].
\eea
The supersymmetry conditions (\ref{411})--(\ref{415}) imply that the
Killing vectors of the $S^2$, together with $\hat{w}$, are Killing
vectors of the full solution that also preserve the flux. Thus supersymmetry implies that the
generic isometry group is $\SU(2)\times U(1)$.
We are unaware of any explicit known solutions in this class.


\section{Supersymmetric $AdS_4$ spacetimes}

\subsection{$AdS_4$ spacetimes from wrapping associative cycles}
These geometries will be dual to three-dimensional SCFTs with minimal
${\cal N}=1$ supersymmetry which, generically, have no $R$-symmetry.

For the associative wrapped brane geometries there is a single overall
transverse direction so that $q=1$. Imposing our $AdS_4$ limit
we have $G'=\SU(3)$ with $e^7=\hat{u}$ and thus
\begin{equation}
   \dd s^2(\mathcal{N}_7) =
      \dd s^2(\mathcal{M}_{\SU(3)})
         + \frac{\lambda^2\dd\rho^2}{4m^2(1-\lambda^3\rho^2)} .
\end{equation}
Defining the orientation
\begin{equation}
   \epsilon = \tfrac{1}{6}J\wedge J \wedge J\wedge\hat{\rho},
\end{equation}
the supersymmetry conditions reduce to
\bea
\label{0}
\dd\left(\lambda^{-3/2}\im\Omega-\rho J\wedge \hat{\rho}\right)
   &=&0,\\
\label{00}
\dd\left(\frac{1}{\lambda\rho}J\wedge J
   +\frac{2}{\lambda^{5/2}\rho^2}\re\Omega\wedge\hat{\rho}
   \right)&=&0,
\eea
where the flux is given by
\bea
F=-\dd\left(\frac{\sss}{\l^{3/2}\r}\re\Omega\right)
   + m\left(\frac{1}{\lambda\rho}J\wedge J
      +\frac{2}{\lambda^{5/2}\rho^2}\re\Omega\wedge\hat{\rho}\right).
\eea
In~\cite{L+S}, Lukas and Saffin analysed the conditions for
supersymmetry for a broad class of $\mathcal{N}=1$ $AdS_4$ spacetimes with $SU(3)$
structure in
M-theory\footnote{The most general ansatz for the Killing
  spinors is given in \cite{cvetic}.}. We have verified that the
three equations for supersymmetry given above imply the conditions
(3.40), (3.43)--(3.47) of~\cite{L+S}. However, our results imply
expressions that are slightly different 
from equations (3.41) and (3.48) of~\cite{L+S}. 

The supersymmetry conditions in this case characterise the most
general minimally supersymmetric $AdS_4$ spacetime in M-theory with
purely magnetic fluxes. To see this, observe that we have derived them
by taking the most general $AdS$ limit of the associative calibration
conditions of \cite{J+D}. These in turn were obtained by setting the
electric flux to zero in the conditions for the most general
minimally supersymmetric $\mathbb{R}^{1,2}$ spacetime in M-theory,
also given in \cite{J+D}. The $AdS_4$ supersymmetry conditions in this case
imply that generically the $AdS$ limit of the associative calibration
conditions has no isometries, which is consistent with lack of $R$
symmetry in the dual SCFT. In section~\ref{solutions} we will recover
the explicit solution of~\cite{acharya} using these results.


\subsection{$AdS_4$ spacetimes from wrapping SLAG cycles in $\SU(3)$
  holonomy}

These geometries are dual to three-dimensional SCFTs with
${\cal N}=2$ supersymmetry, which have a $U(1)$ $R$-symmetry.

For the wrapped brane geometries corresponding to wrapping SLAG
three-cycles there are two overall transverse directions so that
$q=2$. Imposing our $AdS_4$ limit we have $G'=\SU(2)$ in five
dimensions with $e^6=\hat{u}$. Writing $\hat{w}=e^5$ for the one-form
used to define this structure, we have
\begin{equation}
   \dd s^2(\mathcal{N}_7) =
      \dd s^2(\mathcal{M}_{\SU(2)})
      + \hat{w}\otimes\hat{w}
      + \frac{\lambda^2}{4m^2}\left(
            \frac{\dd\rho^2}{1-\lambda^3\rho^2}
            + \rho^2 \dd\phi^2 \right) ,
\end{equation}
where $\phi$ is a coordinate on the $S^1$. Defining the orientation
\begin{equation}
   \epsilon = \tfrac{1}{2}J^1\wedge J^1\wedge\hat{w}
      \wedge\hat{\rho}\wedge\hvol(S^1),
\end{equation}
the supersymmetry conditions reduce to
\bea
\label{101}
\dd[\l^{-1}\sqrt{1-\l^3\r^2}\hat{w}]&=&
   m\l^{-1/2}J^1+m\l\r \hat{w}\wedge\hat{\rho} ,\\
\label{1011}
\dd\left(\l^{-3/2}J^3\wedge\hat{w}-\r J^2\wedge\hat{\rho}\right)
   &=&0,\\\label{10111}
\dd\left(J^2\wedge\hat{w}+\l^{-3/2}\r^{-1}J^3\wedge\hat{\rho}\right)
   &=&0,
\eea
while the flux is given by
\bea\label{515}
F=\frac{1}{\l\r}\hvol(S^1)\wedge
   \dd\left(\l^{-1/2}\sqrt{1-\l^3\r^2}J^3\right).
\eea
The supersymmetry conditions imply that $\der/\der\phi$ is a Killing
vector of the full metric that also preserves the flux. Therefore the SLAG supersymmetry conditions
imply that the $AdS_4$ limit generically has a $U(1)$ isometry group,
as expected from R-symmetry. In section~\ref{solutions} we will
recover the explicit solution of~\cite{gkw} using these results.


\section{Supersymmetric $AdS_5$ spacetimes}

We now turn to the conditions for supersymmetric $AdS_5$ spacetimes
obtained from supersymmetric wrapped-brane geometries. Two cases
remain, corresponding to wrapping K\"{a}hler two-cycles in
$\SU(3)$- and $\SU(2)$-holonomy manifolds. In fact, the first case is
precisely the one considered in~\cite{Gauntlett:2004zh}, where the
reduction to $AdS$ from a wrapped brane geometry was first
discussed. This gave the conditions for the most generic
supersymmetric $AdS_5$ spacetimes in M-theory dual to SCFTs with ${\cal
  N}=1$ supersymmetry. We will not discuss this case any further but
instead we turn directly to the case of wrapping K\"{a}hler two-cycles
in $\SU(2)$-holonomy manifolds, which preserves twice as much
supersymmetry.

\subsection{$AdS_5$ spacetimes from wrapping K\"{a}hler two-cycles in
  $\SU(2)$ holonomy}

These geometries are dual to four-dimensional SCFTs with ${\cal N}=2$
supersymmetry, which have $SU(2)\times U(1)$ $R$-symmetry.

For the K\"{a}hler two-cycle in $\SU(2)$ holonomy wrapped brane
geometries there are three overall transverse directions so that
$q=3$. Imposing our $AdS_5$ limit we find that the $\SU(2)$ structure
is broken to a local identity structure in three dimensions defined by
$(e^1,e^2,e^3)$ with $e^4=\hat{u}$, following the conventions of
appendix~\ref{projs}. We thus have
\begin{equation}
   \dd s^2(\mathcal{N}_6) =
      e^1\otimes e^1 + e^2\otimes e^2 + e^3\otimes e^3
      + \frac{\lambda^2}{4m^2}\left(
            \frac{\dd\rho^2}{1-\lambda^3\rho^2}
            + \rho^2 \dd\Omega_2^2 \right) .
\end{equation}
Defining the orientation
\begin{equation}
   \epsilon = e^{123}\wedge\hat{\rho}\wedge\hvol(S^2),
\end{equation}
the conditions for supersymmetry are
\begin{equation}
\label{65}
\begin{aligned}
   \dd\left(\l^{-1}\sss e^1\right) &=
      m\l^{-1/2}\left(\l^{3/2}\r e^1\wedge\hat{\rho}+e^{23}\right), \\
   \dd\left(\l^{-1}\sss e^2\right) &=
      m\l^{-1/2}\left(\l^{3/2}\r e^2\wedge\hat{\rho}-e^{13}\right), \\
   \dd\left(\frac{\l^{1/2}}{\sss}e^3\right) &=
      - \frac{2m\l}{1-\l^3\r^2}e^{12}
      - \frac{3\l\r}{(1-\l^3\r^2)^{3/2}}\left(
         \partial_{\hat{\rho}}\l e^{12} \right. \\ & \qquad \left. {}
         - \partial_2\l e^1\wedge\hat{\rho}
         + \partial_1\l e^2\wedge\hat{\rho}\right),
\end{aligned}
\end{equation}
and the flux is given by
\bea\label{6.9}
F=-\frac{1}{\l^2\r^2}\hvol(S^2)\wedge\left[
   \dd\left(\l^{1/2}\r\sss e^3\right)
   + 2m\left(\l\r e^{12}+\l^{-1/2}e^3\wedge\hat{\rho}\right)
   \right].
\eea

These backgrounds preserve half of the supersymmetry. These conditions
have in fact already been derived, from a somewhat different perspective, 
by Lin, Lunin and Maldacena (LLM) in~\cite{LLM}.
As we show in detail in Appendix~\ref{app2}, our conditions are indeed
equivalent to those of reference~\cite{LLM}.  Specifically, we find
that for the general solution of the conditions~(\ref{65}), we may
take the metric to be given locally by
\begin{equation}
\begin{aligned}
   \dd s^2(\mathcal{N}_6) &= \frac{\l^2}{4m^2}\left[
      \frac{1}{1-\l^3\r^3}(\dd\r^2+\ee^D\dd x^i\dd x^i)
      + \r^2\dd\Omega_2^2\right] \\ &\qquad\qquad\qquad {}
      + \frac{1-\l^3\r^2}{\lambda m^2}\left(dx^3+V_idx^i\right)^2,
\end{aligned}
\end{equation}
where $i=1,2$, the function $D(\r,x^1,x^2)$ satisfies the Toda
equation
\bea
\left(\partial_{x^1}^2+\partial_{x^2}^2\right)D
   +\partial_{\r}^2\ee^D=0,
\eea
and the function $\l$ and the one-form $V$ are given by
\bea
\l^3&=&-\frac{\partial_{\r}D}{\r(1-\r\partial_{\r}D)},\\
V&=&\tfrac{1}{2}\star_2\dd_2 D,
\eea
where $\dd_2=\dd x^i\,\der_i$. The flux may be read off
from~(\ref{6.9}). Note that here we have not assumed the $SU(2)\times
U(1)$ isometry of these $AdS$ spacetimes, as was done by LLM, but rather we have deduced
it directly from the $AdS$ limit of the supergravity
description of the wrapped brane configuration.


\section{Explicit Solutions}
\label{solutions}

In this section, we will discuss explicit solutions of the
supersymmetry conditions for the $AdS$ geometries we have just
described. We have used two approaches, both of
which reduce the problem to solving ordinary differential
equations.

In the first approach one assumes that the metric on
$\mathcal{M}_{G'}$ is conformal to a standard $G'$-structure metric:
either a special holonomy metric or when $G'=\SU(3)$ a nearly K\"ahler
metric. One then assumes that the conformal factor and the function
$\lambda$ depend only on the coordinate $\rho$.

The second approach is based on the class of known solutions
originally derived using seven-dimensional gauged
supergravity~\cite{M-N,acharya,gkw,gk2} and which describe M5-branes
wrapping a variety of calibrated cycles.  We start by identifying the
relevant structures for these solutions. This will serve as a
highly non-trivial consistency check on our general conditions as well as
elucidating the geometrical structure underlying the solutions. In
addition, this exercise suggests a natural class of
generalisations, again depending only on $\rho$, and we derive the
corresponding ordinary differential equations. In the
case corresponding to wrapping  a K\"{a}hler four-cycle in an
$\SU(3)$-holonomy manifold, we find some new, though singular,
solutions.

We will begin this section by giving a general discussion of the
$G$-structures of the known gauged supergravity $AdS$ solutions. Then
in the following subsections, we discuss the two approaches to finding
more general solutions, in the cases of branes wrapping associative,
co-associative or SLAG cycles, satisfying our general $AdS$
supersymmetry conditions. In particular, we explicitly extract the
$G$-structures underlying the gauged supergravity solutions in each
case. We also include a subsection containing the new singular
solutions for the $AdS_3$ limits in the case of a K\"{a}hler
four-cycle in an $\SU(3)$-holonomy manifold.

\subsection{$G$-structures of gauged supergravity solutions}
\label{gaugedsugra}

A general class of solutions~\cite{M-N,acharya,gkw,gk2} describing
branes wrapping calibrated cycles in the near horizon limit, can be
constructed by first finding $AdS$ solutions in $D=7$ gauged
supergravity and then uplifting to $D=11$. As such, the solutions all
have the form of a warped product of $AdS_{7-d}\times \Sigma_d\times
S^4$, where  $\Sigma_d$ is the cycle that the fivebrane is wrapping
and the four-sphere surrounds the fivebrane. The four-sphere is fibred
over $\Sigma_d$ with the twisting determined by the structure of the
normal bundle of a calibrated cycle in a special holonomy manifold.

More specifically, consider the solution for a fivebrane wrapping a
calibrated $\Sigma_d$ inside a special holonomy manifold. Following the
discussion in~\cite{gkw}, let $p$ denote the number of dimensions
transverse to the fivebrane worldvolume and tangent to the special
holonomy manifold and $q$ the number of dimensions transverse to both
the fivebrane worldvolume and the special holonomy manifold. We thus
have $p+q=5$ and the $SO(5)$ symmetry of a flat fivebrane in flat
space is broken to $SO(p)\times SO(q)$. For the known solutions the
eleven-dimensional metric takes the form
\begin{equation}
\label{81}
\begin{aligned}
   m^2 ds^2 &= \Delta^{-2/5}\left\{ \frac{a_1}{u^2}\left[
          \dd s^2(\bbR^{1,5-d}) + \dd u^2 \right]
        + a_2\dd s^2(\Sigma_d) \right\} \\ &\qquad\qquad{}
     + \Delta^{4/5}\left\{
        \ee^{2q\Lambda} DY^a DY^a
        + \ee^{-2p\Lambda}\dd Y^\alpha\dd Y^\alpha \right\},
\end{aligned}
\end{equation}
where
\bea
\Delta^{-6/5}=e^{-2q\Lambda}(Y^aY^a+e^{10\Lambda}Y^{\alpha}Y^{\alpha}).
\eea
We have written the metric for $AdS_{7-d}$ in Poincar\'e coordinates
which displays the worldvolume of the fivebrane as $\bbR^{1,5-d}\times \Sigma_d$.
The constants $a_1$ and $a_2$ specify the size of the $AdS$ space
and the cycle $\Sigma$. The coordinates $Y^a$, $a=1,\dots p$, and
$Y^\alpha$, $\alpha=1,\dots,q$, with $p+q=5$ parametrise the
four-sphere: $Y^aY^a+Y^\alpha Y^\alpha=1$. We also have
$DY^a=dY^a+B^a{}_bY^b$, where the $SO(p)$ connection $B^a{}_b$ is
determined by the spin connection of the cycle $\Sigma_d$. In
particular, it is determined by the structure of the normal bundle to
the calibrated cycle in the special holonomy
manifold~\cite{M-N}. Furthermore, in the explicit solutions
of~\cite{M-N, acharya,gkw, gk2}   the cycles $\Sigma_d$ are all Einstein,
typically with negative curvature, and satisfy additional conditions
that are discussed in the references.

The examples that are relevant for this paper are those
with vanishing electric four-form components. We have listed the
 values of various quantities for these cases in
Table~\ref{sugrasols}. Note that there is no entry for the K\"ahler 4
in $CY_3$ since there are no known solutions with $AdS_3$ factors.
\begin{table}
\begin{center}
\begin{tabular}{|l||l|l|l|l|l|l|l|}
\hline
&$p$&$q$&$a_1$&$a_2$&$\ee^{10\Lambda}$&$c_1$&$c_2$\\*[2pt]
\hline
SLAG in $CY_3$
   & 3 & 2 &
   $\ee^{8\Lambda}$ & $\ee^{-2\Lambda}$
   & 2 & 1 & 2 \\*[2pt]
K\"ahler 2-cycle in $CY_2$
   & 2 & 3 &
   $2\ee^{2\Lambda}$ & $\ee^{2\Lambda}$
   & 2 & 1 & 2 \\*[2pt]
K\"ahler 2-cycle in $CY_3$
   & 4 & 1 &
   $\frac{9}{4}\ee^{4\Lambda}$ & $\ee^{-6\Lambda}$
   & $\frac{4}{3}$ & $\frac{3}{2}$ & 2 \\*[2pt]
associative in $G_2$-holonomy
   & 4 & 1 &
   $\frac{25}{16}\ee^{4\Lambda}$ & $\ee^{-6\Lambda}$
   & $\frac{8}{5}$ & $\frac{5}{4}$ & 2 \\*[2pt]
co-associative in $G_2$-holonomy
   & 3 & 2 &
   $\frac{4}{9}\ee^{8\Lambda}$ & $\ee^{-2\Lambda}$
   & 3 & $\frac{2}{3}$ & 2 \\*[2pt]
\hline
\end{tabular}
\end{center}
\caption{Examples of wrapped M5-brane solutions}
\label{sugrasols}
\end{table}

To identify the underlying $G$-structure, it is illuminating to change
coordinates from $u$, $Y^a$, $Y^\alpha$ to unconstrained ``cartesian''
coordinates $X^a,X^\alpha$ via
\bea\label{blip}
X^a&=&u^{-c_1}Y^a,\qquad  c_1=\ee^{-2q\Lambda}{\sqrt {a_1}},\nn
X^\alpha&=& u^{-c_2} Y^\alpha,\qquad c_2=\ee^{2p\Lambda}{\sqrt {a_1}},
\eea
to obtain
\begin{equation}
\begin{aligned}
   m^2 ds^2 &= \Delta^{-2/5}\left[
        \frac{a_1}{u^2}\dd s^2(\bbR^{1,5-d})
        + a_2\dd s^2(\Sigma_d) \right] \\ &\qquad\qquad{}
     + \Delta^{4/5}\left[
        \ee^{2q\Lambda}u^{2c_1} DX^a DX^a
        + \ee^{-2p\Lambda}u^{2c_2}\dd X^\alpha\dd X^\alpha \right],
\end{aligned}
\end{equation}
where $DX^a=dX^a+B^a{}_b X^b$. Although we have now obscured the
$AdS_{7-d}$ structure, the world-volume of the fivebrane,
$\bbR^{1,5-d}\times \Sigma_d$, is still manifest and we have
revealed\footnote{The $AdS$ solutions that we are discussing here are
  specific examples of more general solutions still with
  $\bbR^{1,5-d}$ and $\Sigma_d$ factors but a more complicated
  dependence on the coordinate $u$ which describe renormalisation
  group flows ``across dimensions''. The coordinate transformation we
  are describing can be generalised to this more general class of
  solutions. It was first noticed in the context of wrapped membranes
  \cite{gkpw}.} the $\bbR^p\times \bbR^q$ structure of the directions
transverse to the fivebrane either tangent to the original special
holonomy manifold ($\bbR^p$) or transverse to it ($\bbR^q$).

In this form it is straightforward to identify the structure that
corresponds to the supersymmetry conditions that we discussed in
section~\ref{sec:susy}. Clearly $L=\Delta^{2/5}u^2/a_1$. We next note
that for all of the cases considered in section~\ref{sec:susy},
i.e. corresponding to wrapped brane solutions with no electric flux,
we have $c_2=2$. As a consequence, after we rescale $X^\alpha\to
\frac{1}{4}\ee^{5p\Lambda} X^\alpha$ we see that the factor
multiplying the overall transverse directions\footnote{Interestingly
  $c_2\ne 2$ for cases with non-zero electric flux which indicates
  that the factor multiplying the overall transverse directions will
  no longer be $L^2$. The
  most general minimally supersymmetric $AdS_3$ geometry with
  electric flux in M-theory is of this form, as discussed in \cite{J+D}.}
is indeed $L^2$ in agreement with the discussion in
section~\ref{sec:susy}.

To display the rest of the structure in terms of the analysis of the
wrapped-brane backgrounds, it is best to focus on an illustrative
example. Consider the case of wrapping SLAG three-cycles
in manifolds with $\SU(3)$-holonomy. We have $p=3$
corresponding to three directions transverse to the SLAG three-cycle
inside the Calabi--Yau three-fold and $q=2$ corresponding to the two
overall transverse directions. For the solutions given
in~\cite{sezgin,gkw}, the cycle $\Sigma_3$ is hyperbolic space $H_3$
with the standard constant (unit) curvature metric, or a discrete
quotient thereof, which may be compact. For this case the twisting is
such that $B^{a}{}_b=\bar\omega^a{}_b$, the $\SO(3)$ spin connection
of $H_3$. We now let $\bar e^a$ be an orthonormal frame for $\Sigma_3$
and consider the one forms $e^a=\Delta^{-1/5}\sqrt{a_2}m^{-1}\bar e^a$
and $f^a=\Delta^{2/5}\ee^{q\Lambda} u^{c_1}m^{-1} DX^a$.
Then, given the cycle is SLAG, the obvious $SU(3)$ structure is
\begin{equation}
\label{obvst}
\begin{aligned}
   J &= e^a\wedge f^a, \\
   \Omega &= \tfrac{1}{6}\epsilon^{abc}
      (e^a+\ii f^a)(e^b+\ii f^b)(e^c+\ii f^c).
\end{aligned}
\end{equation}
One can check that the SLAG supersymmetry conditions of
section~\ref{SLAG3} are indeed satisfied.

An advantage of displaying the structure at the level of the wrapped
brane solutions is that the structures in the $AdS$ limits are then
easily identified, by carrying out the reduction procedure that we
discussed in section~\ref{sec:red}. Returning to the general case, it
is useful to introduce the following coordinates:
\begin{equation}
\begin{aligned}
   X^a &= u^{-c_1}\cos\tau \tilde{Y}^a,\nn
   X^\alpha &= u^{-c_2}\sin\tau \tilde{Y}^\alpha,
\end{aligned}
\end{equation}
where $\tilde{Y}^a$ parametrise a $(p-1)$-sphere, $\tilde{Y}^a
\tilde{Y}^a=1$, and  $\tilde{Y}^\alpha$ parametrise a $(q-1)$-sphere,
$\tilde{Y}^\alpha \tilde{Y}^\alpha=1$. Obviously this is just
equivalent to $Y^a=\cos\tau \tilde{Y}^a$ and $Y^\alpha=\sin\tau
\tilde{Y}^\alpha$ in~(\ref{blip}). We find that the metric now takes
the form
\begin{equation}
\begin{aligned}
   m^2 \dd s^2 &= \Delta^{-2/5}\left\{
         \frac{a_1}{u^2}\left[
            \dd s^2(\bbR^{1,5-d}) + \dd u^2\right]
         + a_2ds^2(\Sigma_d)
    + \frac{\ee^{2(q-p)\Lambda}}{a_1}\dd\tau^2 \right\} \\
    & \qquad \qquad
    + \Delta^{4/5} \left\{
       \ee^{2q\Lambda}\sin^2\tau\dd\tilde{Y}^\alpha\dd\tilde{Y}^\alpha
       + \ee^{2q\Lambda}\cos^2\tau D\tilde{Y}^a D\tilde{Y}^a
       \right\},
\end{aligned}
\end{equation}
and from the $AdS$ factor we identify $\Delta^{-6/5}=(a_1\lambda)^{-3}=
\ee^{-2q\Lambda}\cos^2\tau+\ee^{2p\Lambda}\sin^2\tau$.
In order to make contact with the coordinates that we used in
section~\ref{sec:red}, we introduce
\begin{equation}
   \rho=2a_1\ee^{-p\Lambda}\sin\tau.
\end{equation}
We then find, for the cases with no electric flux that we are
focussing on in this paper, that the metric
becomes
\begin{equation}
\begin{aligned}
   m^2 \dd s^2 &= \lambda^{-1}\left[
      \dd s^2(AdS_{7-d}) + \frac{a_2}{a_1}\dd s^2(\Sigma_d)
      + \frac{\ee^{20\Lambda}}{4}(1-\l^3\r^2)D\tilde{Y}^a D\tilde{Y}^a
      \right]
      \\ & \qquad \qquad \qquad \qquad
      +\frac{\lambda^2}{4}\left(
         \frac{\dd\rho^2}{1-\lambda^3\rho^2}
         + \rho^2 \dd\tilde{Y}^\alpha \dd\tilde{Y}^\alpha \right).
\end{aligned}
\end{equation}
This agrees with the general form~\eqref{N-metric} if we identify
\begin{equation}
   \dd s^2(\mathcal{M}_{G'})
      = \frac{a_2}{a_1\lambda m^2}\dd s^2(\Sigma_d)
      + \frac{\ee^{20\Lambda}}{4\lambda m^2}
         (1-\l^3\r^2)D\tilde{Y}^a D\tilde{Y}^a .
\end{equation}

To identify the $G$-structure in  the $AdS$ limit, returning to the
$\tau$ coordinate, we define the one-form
\begin{equation}
\begin{aligned}
   \hat{u} &= \frac{\Delta^{2/5}\ee^{q\Lambda}}{m}\left(
         c_1\cos\tau\frac{\dd u}{u} + \sin\tau\dd\theta \right) \\
      &= \l^{-1/2}\sss\dd r+\frac{\l^{5/2}\r\dd\r}{2m\sss},
\end{aligned}
\end{equation}
where we have defined $r=m^{-1}\log u$. This matches the unit
one-form~\eqref{r-decomp} introduced in section~\ref{sec:red} which
gives the component of the $AdS$ radial direction in the space
$\mathcal{M}_G$. We then have
\bea
   \frac{\Delta^{2/5} \ee^{q\Lambda}}{m}u^{c_1}DX^a
      = - \tilde{Y}^a\hat{u}
         + \frac{\Delta^{2/5}\ee^{q\Lambda}}{m}\cos\tau D\tilde{Y}^a.
\eea
For the SLAG three-cycle case the two-form $J$ of the $\SU(3)$
structure introduced above~(\ref{obvst}) now becomes
\be
J=J^1+\hat{w}\wedge \hat{u},
\ee
where
\bea
\hat{w}&=&-\frac{\sqrt {a_2}\Delta^{-1/5}}{m}\tilde{Y}^a\bar{e}^a,\nn
J^1&=&\frac{\sqrt{a_2}\Delta^{1/5}\ee^{q\Lambda}}{m^2}
   \cos\tau\, \bar e^a \wedge D\tilde{Y}^a.
\eea
Furthermore analysing the expression for $\Omega$, using its
decomposition under $SU(2)$ as given in appendix~\ref{projs}, implies
that
\bea
J^2 &=& - \frac{\sqrt{a_2}\Delta^{1/5}\ee^{q\Lambda}\cos\tau}{m^2}
     \epsilon^{abc}
        \tilde{Y}^a D\tilde{Y}^b\wedge \bar e^c , \nn
J^3 &=& \frac{\Delta^{4/5}\ee^{2q\Lambda}\cos^2\tau}{2m^2}
    \epsilon^{abc}
        \tilde{Y}^a D\tilde{Y}^b \wedge D\tilde{Y}^c
    - \frac{a_2\Delta^{-2/5}}{2m^2}\epsilon^{abc}
        \tilde{Y}^a \bar e^b\wedge \bar e^c .
\eea

The structures of the $AdS$ solutions in the other
cases may be similarly identified, both at the level of the
wrapped brane structures and at the level of the structures in the
$AdS$ limits. In the next subsections we will use this to motivate
some ans\"{a}tze for solving the $AdS$ supersymmetry
conditions which include the explicit uplifted gauged supergravity
solutions as special cases.

\subsection{$AdS_3$ solutions from wrapping co-associative cycles}

In this subsection, we will discuss three simple choices of the $\SU(3)$
structures which arise from the $AdS_3$ limit of the co-associative
wrapped-brane spacetime, and which lead to simple solutions. Two are
based on the first approach where we assume that the metric on
$\mathcal{M}_{SU(3)}$ is conformal to a standard geometry and the third
follows from the structure of the known gauged supergravity solution.

\paragraph{$AdS_3\times S^2\times CY_3$ solutions}
The simplest family of solutions in the co-associative class is
obtained by taking $\l=\text{constant}$ (up to an overall rescaling of
the metric, we may choose $\l=1$), and taking $\mathcal{M}_{SU(3)}$ to be a
Calabi--Yau, independent of the coordinates $\r$ and $\phi$. Then it
is immediately clear that equations~(\ref{44}) and~(\ref{45}) are
satisfied, and the metric becomes the direct product $AdS_3\times
S^2\times CY_3$. In fact, these are precisely the
solutions~\eqref{CYsol} we found in the K\"{a}hler four-cycle in
$\SU(3)$ holonomy class, when the $AdS$ radial direction lay entirely
in the overall transverse space. It is entirely consistent that they
also arise here, since wrapping a K\"{a}hler four-cycle is a special
case of wrapping a co-associative cycle in a $G_2$ holonomy
manifold. Specifically we can write the $G_2$ structure metric
in~\eqref{co-ass-g} as
\bea
   \dd s^2(\mathcal{M}_{G_2})
      = \dd s^2(\mathcal{M}_{\SU(3)}) + L^2 \dd y_3^2,
\eea
giving the same form as for the K\"ahler four-cycle wrapped brane
geometry~\eqref{SLAG-g}. Choosing the $AdS$ radial direction to lie
solely in the overall transverse space of the latter geometry thus
corresponds to it lying partly in the overall transverse space and
partly in $\mathcal{M}_{G_2}$ when viewed as a co-associative
wrapped-brane geometry. From this perspective, the $AdS_3\times
S^2\times CY_3$ solution to the K\"ahler four-cycle class is a special
case of the co-associative $AdS$ geometries, preserving twice as many
supersymmetries.

\paragraph{Nearly-K\"ahler solutions}
A second family of solutions is obtained by assuming the metric on
$\mathcal{M}_{SU(3)}$ is not Calabi--Yau but is conformal to a nearly
K\"ahler geometry. One takes
\bea
\l&=&\l(\r),\nn
\dd s^2(\mathcal{M}_{SU(3)})&=&g^2(\r)\,\dd\tilde{s}^2(\mathcal{M}_{SU(3)}),
\eea
where the nearly K\"{a}hler (NK) metric $\dd\tilde{s}^2$ is independent of
$\r$ and $\phi$. The NK structure implies that the rescaled local
$SU(3)$ structure given by $\tilde{J}=g^{-2}J$ and
$\tilde{\Omega}=g^{-3}\Omega$ satisfies
\bea
\dd\im\tilde{\Omega} & =&0, \nn
\dd\re\tilde{\Omega} &=& c\tilde{J}\wedge\tilde{J},\nn\label{72}
\dd\tilde{J} &=&\frac{3}{2}c\im\tilde{\Omega},
\eea
with $c$ a constant. In this case, equations~(\ref{44}) and~(\ref{45})
reduce to
\bea
\frac{\dd}{\dd\r}(g^3)&=&-\frac{3cg^2}{4\l^{1/2}\r m\sss},\\
\frac{\dd}{\dd\r}\Big(\frac{g^4}{\l}\Big)&=&-\frac{c\l^{3/2}\r g^3}{m\sss}.
\eea
Unfortunately, we have not found any explicit solutions of
these equations. It may of course turn out to be the case that the
general solution of these equations is singular, or that the metric of
the general solution has the wrong signature (owing to the presence of
the $1-\l^3\r^2$ term).

\paragraph{Gauged supergravity inspired solutions}
Now let us recover and generalise the known gauged supergravity
solution given in~\cite{gkw}. We make the metric ansatz
\bea
\l&=&\l(\r),\nn
\dd s^2(\mathcal{M}_{SU(3)})&=&g^2(\r)\dd s^2(\Sigma_4)
   + f^2(\r)D\tilde{Y}^aD\tilde{Y}^a,
\eea
where the $\tilde{Y}^a$ are constrained coordinates on an $S^2$,
$\tilde{Y}^a\tilde{Y}^a=1$, and if $J^a$, $a=1,2,3$ are a triplet of
self-dual two-forms on $\Sigma_4$, taking the standard form (\ref{A13}), then
\bea
D\tilde{Y}^a
   =\dd\tilde{Y}^a-\frac{1}{2}\epsilon^{abc}\tilde{Y}^b\omega_{ij}J^{cij},
\eea
with $i,j=1,\dots,4$. Then we make the following ansatz for the $SU(3)$
structure:
\bea
J &= & g^2\tilde{Y}^aJ^a
+\frac{1}{2}f^2\e^{abc}\tilde{Y}^aD\tilde{Y}^b\wedge D\tilde{Y}^c,\\
\mbox{Im}\Omega&=&g^2fD\tilde{Y}^a\wedge J^a,\\
\mbox{Re}\Omega&=&g^2f\e^{abc}\tilde{Y}^aD\tilde{Y}^b\wedge J^c.
\eea

We begin by noting that $d(\tilde{Y}^a J^a)=D\tilde{Y}^a\wedge J^a$
since it can be shown that $DJ^a=0$.
In addition we have $D^2
\tilde{Y}^a=(1/4)\epsilon^{abc}J^{bij}R_{ijkl}e^k\wedge e^l
\tilde{Y}^c$ where $R_{ijkl}$ is the Riemann tensor of the metric on
$\Sigma_4$. To demonstrate these facts it is useful to introduce a
basis of anti-self-dual tensors $K^a_{ij}$ satisfying
$K^{a}_{ij}K^{bj}_{\;\;\;k}=-\delta^{ab}\delta_{ik}+\epsilon^{abc}K^c_{ik}$
and to note that $K^{a}_{ij}J^{bj}_{\;\;\;k}$ is a symmetric traceless
tensor for each $a,b$. Furthermore, it is helpful to observe that
$(1/2)J^{aij}J^a{}_{kl}$ is a projector onto self-dual tensors. It
will also be useful to recall that in four dimensions the Riemann
tensor can be decomposed as follows:
\bea
R_{ijkl}=C_{ijkl}+\delta_{i[k}\hat{R}_{l]j}-\delta_{j[k}\hat{R}_{l]i}
+\frac{R}{6}\delta_{i[k}\delta_{l]j},
\eea
where $C_{ijkl}$ is the Weyl tensor, $\hat{R}_{ij}$
denotes the traceless part of the Ricci tensor,
\bea
R_{ij}=\hat{R}_{ij}+\frac{R}{4}\delta_{ij},
\eea
with $R$ the scalar Ricci curvature. In addition, we observe that the
Weyl tensor may be expressed as
\bea
C_{ijkl}=A^{ab}J^a_{ij}J^b_{kl}+B^{ab}K^a_{ij}K^b_{kl},
\eea
for some symmetric traceless $A^{ab}$, $B^{ab}$.

Using these results, if $\tilde{d}$ denotes the exterior derivative
restricted to $\mathcal{M}_{SU(3)}$, we then find
\bea
\tilde{d}J&=&\Big(g^2+\frac{Rf^2}{12}\Big)D\tilde{Y}^a\wedge
J^a+\frac{f^2}{4}D\tilde{Y}^a\wedge J^{aij}C_{ijkl}e^k\wedge
e^l\nn&&+\frac{f^2}{2}D\tilde{Y}^a\wedge J^{aj}_k\hat{R}_{lj}e^k\wedge e^l,
\eea
where $C_{ijkl}$, $\hat{R}_{ij}$ and $R$ denote respectively the Weyl
tensor, the traceless part of the Ricci tensor and the Ricci scalar of
$\Sigma_4$. From equation (\ref{44}) we find that we must have
\bea
J^{aij}C_{ijkl}=\hat{R}_{ij}=0,
\eea
so $\Sigma_4$ must be conformally half-flat Einstein. Furthermore, it
is readily verified that $\tilde{d}\mbox{Im}\Omega=0$. Equation
(\ref{44}) also gives the condition
\bea\label{46}
\frac{d}{d\r}(g^2f)=-\frac{1}{2m\l^{1/2}\r\sss}\Big(g^2+\frac{Rf^2}{12}\Big).
\eea
Given the conditions on the curvature of $\Sigma_4$, we find that
\bea
\tilde{d}\mbox{Re}\Omega=
\frac{g^2fR}{3}\mbox{Vol}(\Sigma_4)+g^2f\tilde{Y}^dJ^d\wedge
\e^{abc}\tilde{Y}^aD\tilde{Y}^b\wedge D\tilde{Y}^c.
\eea
In deriving the second term we found the following identity useful:
\be
[\delta^{dc}-\tilde{Y}^d\tilde{Y}^c]
\epsilon^{cab}D\tilde{Y}^a\wedge D\tilde{Y}^b=0.
\ee
Then noting that $\tilde d( J\wedge J)=0$, (\ref{45}) gives the conditions
\bea\label{714}
\frac{d}{d\r}\Big(\frac{g^4}{\l}\Big)&=&-\frac{g^2f\l^{3/2}\r
  R}{6m\sss},\\\label{715}
\frac{d}{d\r}\Big(\frac{g^2f^2}{\l}\Big)&=&-\frac{g^2f\l^{3/2}\r}{m\sss}.
\eea

This pair of equations, together with the curvature conditions on
$\Sigma_4$ and~(\ref{46}), are exhaustive for our ansatz. We observe
that choosing $f=g$, $R=6$, equations (\ref{46}), (\ref{714}) and
(\ref{715}) reduce to the equations for the nearly K\"{a}hler
family discussed above, since then our ansatz together with the curvature
conditions on $\Sigma_4$ implies that $\mathcal{M}_{SU(3)}$ is nearly
K\"{a}hler.

We have not found the general solution of (\ref{46}),
(\ref{714}) and (\ref{715}). However, it is
readily verified that a particular solution is given by
\bea
R&=&-4,\nn
g^2&=&\frac{3}{4\l m^2},\nn
f^2&=&\frac{9(1-\l^3\r^2)}{4\l m^2}\nn
\l^3&=&\frac{3}{2(\r^2+\a)},
\eea
for some constant $\a$, which is essentially irrelevant as it must be
positive for the metric to have the correct signature, and it may then
be
absorbed into an overall scale in the metric by rescaling $\r$. It may
be verified that choosing $\a=32/27$ and defining constrained coordinates
on an $S^4$, ${Y}^a{Y}^a+Y^{\a}Y^{\a}=1$, $\a=4,5$,
according to
\bea
Y^a&=&\sqrt{1-\frac{27}{64}\r^2}\tilde{Y}^a,\\
Y^4&=&\sqrt{\frac{27}{64}}\r\sin\phi,\\
Y^5&=&\sqrt{\frac{27}{64}}\r\cos\phi,
\eea
we obtain precisely the metric (\ref{81}), in the co-associative case,
and hence the solution of~\cite{gkw}.

\subsection{$AdS_3$ spacetimes from wrapping K\"{a}hler four-cycles}

Now we turn to the construction of two distinct (singular) families of
$AdS_3$ spacetimes from K\"{a}hler-4 in $CY_3$ wrapped brane spacetimes, where the
$AdS$ radial direction lies partly in the $CY_3$ and partly in the
overall transverse space. To the best of our knowledge, these are the
first examples of this class of solutions to be constructed. (Recall
that there are no gauged supergravity solutions in this class and that
the $AdS_3\times S^2\times CY_3$ solution has the radial direction
just in the overall transverse directions).

For the first family, we make the following ansatz
\bea
\l&=&\l(\r),\\
\hat{w}&=&f(\r)\tilde{w}(x),\\
ds^2(\mathcal{M}_{SU(2)})&=&g^2(\r)\dd\tilde{s}^2(\mathcal{M}_{SU(2)})
\eea
where $\dd\tilde{s}^2$ is a $\rho$-independent metric of $SU(2)$ holonomy. Equations
(\ref{411}) and (\ref{412}) imply that
\bea
g^2=\frac{\l^{1/2}}{\sss}.
\eea
Equations (\ref{414}) and (\ref{415}) imply that
\bea
\hat{w}&=&\frac{\sss}{\l^{1/2}}(d\psi +A(x)),\\
dA&=&2m\alpha \tilde{J}^1+K^{(-)},
\eea
where $\alpha=\mbox{constant}$, $A$ is a one-form on $\mathcal{M}_{SU(2)}$,
$\tilde{J}^1=g^{-2}J^1$, and $K^{(-)}$ denotes an arbitrary
anti-self-dual two-form on $\mathcal{M}_{SU(2)}$. Equation (\ref{413}) then
becomes
\bea
\frac{d}{d\r}\Big(\frac{\l^{3/2}\r}{\sss}\Big)\tilde{J}^1
+\alpha\tilde{J}^1+\frac{1}{2m}K^{(-)}=0,
\eea
whence
\bea
K^{(-)}&=&0,\\
\l^3&=&\frac{(\beta-\alpha\r)^2}{\r^2[1+(\beta-\alpha\r)^2]},
\eea
for some constant $\beta$. We have now solved all the supersymmetry
conditions, but it is readily verified that the resulting solution is
singular for all values of $\alpha$ and $\beta$.

A second family of singular solutions may be obtained as follows. We
make the ansatz
\bea
\l&=&\l(\r),\\e^a&=&g(\r)dx^a,\;\;a=1,2,\\
e^p&=&h(\r)dx^p\;\;p=3,4,\\
\hat{w}&=&f(\r)(d\psi+A(x)),\\
dA&=&2m\alpha dx^1\wedge dx^2+2m\beta dx^3\wedge dx^4,
\eea
where $A$ is a one-form on $\mathcal{M}_{SU(2)}$, and $\alpha$ and $\beta$ are
constants. Then equations (\ref{411}) and (\ref{412}) imply that
\bea
gh=\frac{\l^{1/2}}{\sss}.
\eea
Equations (\ref{414}) and (\ref{415}) give
\bea
f=\frac{\sss}{\l^{1/2}}.
\eea
The final condition for supersymmetry we must solve is (\ref{413}),
which reads
\bea
\frac{d}{d\r}(\l\r g^2)&=&-\alpha,\\
\frac{d}{d\r}\Big(\frac{\l^2\r}{(1-\l^3\r^2)g^2}\Big)&=&-\beta.
\eea
Therefore
\bea
g^2&=&\frac{\gamma-\alpha\r}{\l\r},\\\l^3&=&\frac{(\gamma-\alpha\r)(\delta-\beta\r)}{\r^2[1+(\gamma-\alpha\r)(\delta-\beta\r)]},
\eea
for some constants $\gamma,\delta$. It is readily verified that these
solutions are singular for all values
of $\alpha,\beta,\gamma,\delta$.

\subsection{$AdS_4$ spacetimes from wrapping associative cycles}
In this subsection, we will examine two distinct families of $AdS_4$
spacetimes arising from associative calibrations. As in the
co-associative case, one of these families involves a nearly
K\"{a}hler manifold, and we reduce the problem in this case to a pair of 
first-order ODEs, though we have not been able to find any explicit
solutions. For the second family, we show that the problem reduces to
the determination of three functions satisfying three first-order
ODEs, which we show are satisfied by the explicit solution first constructed in
gauged supergravity in \cite{acharya}.

\paragraph{$NK_6$ solutions}
To obtain the equations governing this family of solutions, we make the
same metric ansatz as in the co-associative case, namely
\bea
\l&=&\l(\r),\nn
\dd s^2(\mathcal{M}_{SU(3)})&=&g^2(\r)\dd \tilde{s}^2(\mathcal{M}_{SU(3)}),
\eea
with $\dd \tilde{s}$ a $\rho$-independent nearly K\"{a}hler metric on $\mathcal{M}_{SU(3)}$, so that it admits an $SU(3)$ structure
satisfying (\ref{72}). Then equations (\ref{0}) and (\ref{00}) become
\bea
\frac{d}{d\r}\Big(\frac{g^3}{\l^{3/2}}\Big)&=&-\frac{3c\l\r
  g^2}{4m\sss},\\
\frac{d}{d\r}\Big(\frac{g^4}{\l\r}\Big)&=&-\frac{cg^3}{\l^{3/2}\r^2m\sss}.
\eea
Unfortunately, we have not found any explicit solutions of
these equations. This family of solutions has
also been discussed in \cite{L+S}.

\paragraph{Gauged supergravity inspired solutions}
Let us now recover the explicit associative $AdS_4$ solution
constructed in gauged supergravity in \cite{acharya}. We make the
metric ansatz
\bea
\l&=&\l(\r),\nn
ds^2(\mathcal{M}_{SU(3)})&=&f^2(\r)\mu^a\mu^a+g^2(\r)ds^2(\Sigma_3),
\eea
where $a=1,2,3$, $e^a$ are a basis for $\Sigma_3$, and
\bea
\mu^a=\sigma^a-\frac{1}{2}\e^{abc}\omega_{bc},
\eea
where the $\sigma^a$ are left-invariant one-forms on an $S^3$,
$d\sigma^a=\frac{1}{2}\e^{abc}\sigma^b\wedge \sigma^c$. We make the
following ansatz for the $SU(3)$ structure
\bea
J&=&fg\mu^ae^a,\\
\mbox{Im}\Omega&=&\frac{1}{2}f^2g\e^{abc}e^a\wedge
\mu^b\wedge\mu^c-\frac{1}{6}g^3\e^{abc}e^a\wedge e^b \wedge e^c,\\
\mbox{Re}\Omega&=&
\frac{1}{6}f^3\e^{abc}\mu^a\wedge\mu^b\wedge\mu^c-\frac{1}{2}fg^2\e^{abc}\mu^a\wedge
e^b\wedge e^c.
\eea
Denoting the exterior derivative restricted to $\mathcal{M}_{SU(3)}$ by
$\tilde{d}$, we find
\bea
\tilde{d}J&=&\frac{1}{2}fg\e^{abc}\mu^a\wedge\mu^b\wedge
e^c-\frac{1}{12}fgR\e^{abc}e^a\wedge e^b\wedge e^c,\\
\tilde{d}\mbox{Im}\Omega&=&0,\\
\tilde{d}\mbox{Re}\Omega&=&-f^3\hat{R}_{bc}e^a\wedge e^b\wedge
\mu^a\wedge \mu^c-\Big(\frac{Rf^3}{12}+\frac{1}{2}fg^2\Big)e^a\wedge
e^b\wedge\mu^a\wedge\mu^b,
\eea
where $\hat{R}_{ab}$ and $R$ are respectively the traceless part of the
Ricci tensor and the Ricci scalar of $\Sigma_3$. In three dimensions,
\bea
R_{abcd}=2(\delta_{a[c}\hat{R}_{d]b}-\delta_{b[c}\hat{R}_{d]a})+\frac{R}{3}\delta_{a[c}\delta_{d]b}.
\eea

We observe that if one sets $f=g$ and $R=2$, one then gets a special case 
of the $NK_6$ family of solutions discussed above.
In general, (\ref{0}) and (\ref{00}) imply that
$\hat{R}_{ab}=0$, $R$ is constant and that
\bea
\frac{d}{d\r}\Big(\frac{f^2g}{\l^{3/2}}\Big)&=&-\frac{\l\r
  fg}{2m\sss},\\
\frac{d}{d\r}\Big(\frac{g^3}{\l^{3/2}}\Big)&=&-\frac{\l\r
  fgR}{4m\sss},\\
\frac{d}{d\r}\Big(\frac{f^2g^2}{\l\r}\Big)&=&-\frac{1}{\l^{3/2}\r^2
  m\sss}\Big(\frac{Rf^3}{12}+\frac{1}{2}fg^2\Big).
\eea

We have not found the general solution of these equations.
However, it is readily verified that a particular solution is given by
\bea
R&=&-6,\\
f^2&=&\frac{4}{25\l m^2}(1-\l^3\r^2),\\
g^2&=&\frac{4}{5\l m^2},\\
\l^3&=&\frac{8}{5}\frac{1}{(1+\frac{3}{5}\r^2)}.
\eea
If we define a new coordinate $\theta$ such that
\bea
\r=\sin\theta,
\eea
then up to an overall constant scale the metric is given by
\bea
ds^2&=&4\Delta^{1/3}X^8\Big[ds^2(AdS_4)+\frac{4}{5}ds^2(\Sigma_3)\Big]+X^3\Delta^{1/3}d\theta^2+\frac{1}{4}\Delta^{-2/3}X^{-1}\cos^2\theta\mu^a\mu^a,\nn&&
\eea
where
\bea
X&=&\Big(\frac{5}{8}\Big)^{1/5},\nn
\Delta&=&X^{-4}\sin^2\theta+X\cos^2\theta.
\eea
This is exactly the eleven dimensional lift of the gauged
supergravity solution given in section (3.2) of \cite{acharya}, in its
original form (setting $h=1$ in \cite{acharya}).
This may of course also be set in the form (\ref{81}).

\subsection{$AdS_4$ spacetimes from wrapping SLAG cycles}
In this subsection, we will discuss a gauged supergravity inspired ansatz
for the $AdS_4$ spacetime arising from M5-branes wrapping a SLAG three-cycle.
In fact we will be able to solve all of the resulting equations and we will find
that the gauged supergravity solution discussed in section 9.1 is the only regular one.

We make the following metric ansatz:
\bea
\l&=&\l(\r),\\
\hat{w}\otimes\hat{w}+ds^2(\mathcal{M}_{SU(2)})&=&f^2(\rho)D\tilde{Y}^aD\tilde{Y}^a+g^2(\r)ds^2(\Sigma_3),
\eea
where $a=1,...,3$, and again the $\tilde{Y}^a$ are constrained coordinates on an
$S^2$, satisfying $\tilde{Y}^a\tilde{Y}^a=1$. We define
\bea
D\tilde{Y}^a=d\tilde{Y}^a+\omega^a_{\;\;\;b}\tilde{Y}^b,
\eea
where $\omega_{ab}$ is the spin connection of $\Sigma_3$. We let $e^a$
denote a basis for $\Sigma_3$, which we assume not to depend on the
$\tilde{Y}^a$. Then we make the following ansatz for the structure:
\bea
\hat{w}&=&g\tilde{Y}^ae^a,\\
J^1&=&fgD\tilde{Y}^a\wedge e^a,\\
J^2&=&fg\epsilon^{abc}\tilde{Y}^aD\tilde{Y}^b\wedge e^c,\\
J^3&=&\frac{1}{2}\epsilon^{abc}[f^2\tilde{Y}^aD\tilde{Y}^b\wedge D\tilde{Y}^c-g^2\tilde{Y}^ae^b\wedge
e^c].
\eea
Note that we have flipped the signs of $\hat{w}$, $J^1$ and $J^2$ with
respect to their definitions in subsection \ref{gaugedsugra}.
We now insert this ansatz into (\ref{101}), (\ref{1011}) and
(\ref{10111}). Equation (\ref{101}) immediately yields
\bea
f&=&\frac{\sss}{\l^{1/2}m},\\\label{525}
\frac{d}{d\r}\Big(\frac{\sss g}{\l}\Big)&=&-\frac{\l^2\r g}{2\sss}.
\eea
The analysis of (\ref{1011}) and (\ref{10111}) is significantly more
complicated, and it is helpful to use an equivalent form of these
equations:
\bea\label{524}
d\Big(\frac{1}{\l^{1/2}\sss}J^3\Big)\wedge
\hat{w}&=&-\frac{m\l^{3/2}\r}{1-\l^3\r^2}J^3\wedge
\hat{w}\wedge\hat{\r}\nn&&+\r d\Big(\frac{\l}{\sss}J^2\Big)\wedge\hat{\r},\\\label{526}
d\Big(\frac{\l}{\sss}J^2\Big)\wedge
\hat{w}&=&-\frac{m\l^{3}\r}{1-\l^3\r^2}J^2\wedge
\hat{w}\wedge\hat{\r}\nn&&-\r^{-1}
d\Big(\frac{1}{\l^{1/2}\sss}J^3\Big)\wedge\hat{\r}.
\eea
We also note that $D^2\tilde{Y}^a=\frac{1}{2}R^a_{\;\;bcd}\tilde{Y}^be^c\wedge
e^d$. Now consider (\ref{526}). It is straightforward to show that
\bea\label{bvc}
&&d\Big(\frac{\l}{\sss}J^2\Big)\wedge\hat{w}=-2m\frac{d}{d\r}\log(\l^{1/2}g)J^2\wedge \hat{w}\wedge
\hat{\r}\nn&&+\Big[(\delta^{ab}-\tilde{Y}^a\tilde{Y}^b)\omega_{abc}d\tilde{Y}^c+\epsilon^{abc}\epsilon^{def}\tilde{Y}^a\tilde{Y}^ed\tilde{Y}^b\omega_{fcd}\Big]\wedge
Vol(\Sigma_3),
\eea
where
\bea
\epsilon^{abc}Vol(\Sigma_3)=e^a\wedge e^b\wedge e^c.
\eea
After some manipulation, the term is square brackets in (\ref{bvc})
may be shown to vanish. Next, we find that
\bea
-\frac{1}{\r}d\Big(\frac{1}{\l^{1/2}\sss}J^3\Big)\wedge
\hat{\r}&=&\Big[-\frac{\sss}{2\l^{3/2}\r m^2}d(\epsilon^{abc}\tilde{Y}^aD\tilde{Y}^b\wedge
D\tilde{Y}^c)\nn&&+\frac{g^2}{2\l^{1/2}\r\sss}\epsilon^{abc}D\tilde{Y}^a\wedge e^b\wedge
e^c\Big]\wedge \hat{\r}.\nn&&
\eea
Observing that
\bea
J^2\wedge \hat{w}=\frac{fg^2}{2}\epsilon^{abc}D\tilde{Y}^a\wedge e^b\wedge e^c,
\eea
equation (\ref{526}) becomes
\bea\label{530}
\Big(2m\frac{d}{d\r}\log(\l^{1/2}g)+\frac{m}{\r}\Big)J^2\wedge
\hat{w}-\frac{\sss}{2\l^{3/2}\r m^2}d(\epsilon^{abc}\tilde{Y}^aD\tilde{Y}^b\wedge D\tilde{Y}^c)=0.
\eea
It may be shown that
\bea
d(\epsilon^{abc}\tilde{Y}^aD\tilde{Y}^b\wedge
D\tilde{Y}^c)=\epsilon^{abc}\tilde{Y}^a\tilde{Y}^dR_{bdef}D\tilde{Y}^c\wedge e^e\wedge e^f,
\eea
where $R_{abcd}$ are the components of the Riemann tensor of
$\Sigma_3$. Then (\ref{530}) becomes
\bea
&&\Big(2m\frac{d}{d\r}\log(\l^{1/2}g)+\frac{m}{\r}+\frac{R}{6m\l\r
  g^2}\Big)J^2\wedge \hat{w}+\frac{1}{g\l\r m}J^2\wedge
(\tilde{Y}^a\hat{R}_{ab}e^b)\nn&&+\frac{\sss}{g\l^{3/2}\r
  m^2}\epsilon^{abc}\tilde{Y}^aD\tilde{Y}^b\wedge(\hat{R}_{cd}e^d)\wedge \hat{w}=0.
\eea
Taking the wedge product of this equation with $\hat{w}$, we discover that
\bea
\tilde{Y}^a\hat{R}_{ab}d\tilde{Y}^b=0,
\eea
and then taking the wedge product with $J^2$ we find that
\bea
\epsilon^{abc}\tilde{Y}^aD\tilde{Y}^b\wedge(\hat{R}_{cd}e^d)\wedge \hat{w}=0.
\eea
Therefore
\bea
\hat{R}_{ab}=0,
\eea
and thus $\Sigma_3$ is required to be Einstein, so it must be either
$H^3$, $S^3$, or some quotient thereof. The remaining condition
contained in (\ref{526}) is
\bea\label{535}
2\frac{d}{d\r}\log(\l^{1/2}g)+\frac{1}{\r}+\frac{R}{6m^2\l\r g^2}=0.
\eea
We may now obtain the general solution of (\ref{525}), (\ref{535}),
and the full set of conditions we have derived hitherto may be
summarised as follows.
\bea
&&\Sigma_3\;\;\mbox{is Einstein},\\
&&f=\frac{\sss}{m\l^{1/2}},\\
&&g=\frac{1}{m\l^{1/2}}\Big(\frac{\alpha}{\r}-\frac{R}{6}\Big)^{1/2},\\
&&\l^3=\frac{12\alpha-2R\r}{12\beta\r-R\r^3},
\eea
for some constants $\alpha,\beta$.

It remains to impose equation (\ref{524}). After a similarly lengthy
analysis, it may be shown that the only additional condition implied
by this equation is
\bea
\alpha=0.
\eea
Then for a real non-singular metric we must choose $\Sigma_3$ such
that $R<0$. The constant $\beta$ is essentially irrelevant; in order
for the metric  to have the correct signature, we must take $\beta<0$,
and then by a
constant rescaling of $\r$ it may be fixed to any particular value, up
to an overall rescaling of the metric. Upon normalising $R=-3$,
choosing $\beta$ such that
\bea
\l^3 =\frac{1}{4(1+\r^2/8)},
\eea
and defining constrained coordinates on an
$S^4$, $Y^a$, $Y^{\alpha}$, $\alpha=4,5$,
$Y^aY^a+Y^{\alpha}Y^{\alpha}=1$, such that
\bea
Y^a&=&\sqrt{1-\frac{\r^2}{8}}\tilde{Y}^a,\\
Y^{4}&=&\frac{1}{2\sqrt{2}}\r\sin\phi,\\
Y^5&=&\frac{1}{2\sqrt{2}}\r\cos\phi,
\eea
we obtain the metric (\ref{81}), in the SLAG case. This is the eleven
dimensional lift of the seven-dimensional solution originally found in
\cite{sezgin}.

\section{Conclusions}
\label{concl}

In this paper we have given a general classification of supersymmetric
geometries with $AdS_{d+2}$ factors in M-theory in terms of $G$-structures. We
have shown that the geometries can be obtained from an interesting class
of spacetimes containing $\bbR^{1,d}$ factors and preserving
algebraically the same set of Killing spinors as a probe M5-brane
wrapping a calibrated cycle in a special holonomy manifold. We have also
characterised this latter class of supersymmetric ``wrapped-brane''
spacetimes in terms of the corresponding $G$-structures.

The technique we have used for characterising the $AdS$ geometries, by
viewing them as special cases of Minkowski geometries of one dimension
less, has allowed us to investigate numerous distinct classes in a way
that is technically reasonably straightforward. 
Of course, the trade-off for this simplification is the loss of the guarantee of
complete generality. However, in the case of $AdS_5$, it was shown in~\cite{Gauntlett:2004zh} that
this approach does in fact lead to the most general supersymmetric
$AdS_5$ geometries dual to ${\cal N}=1$ SCFTs. The work of \cite{J+D},
together with the results here, shows that this is also true for
 $AdS_4$ geometries with purely magnetic flux dual to $\mathcal{N}=1$
 SCFTs. The work of \cite{LLM}, combined with our results,
 strongly suggests that it is true for $AdS_5$ geometries dual to
 $\mathcal{N}=2$ SCFTs, and we strongly suspect that it is also true
 for $AdS_4$ geometries with purely magnetic flux dual to $\mathcal{N}=2$
 SCFTs. For $AdS_3$ geometries with vanishing electric flux that are  dual to
${\cal N}=(2,0)$ supersymmetry it may also be true, though it may
be, for example, that $AdS_3$ geometries arising from M5 branes wrapping
K\"{a}hler four-cycles in Calabi-Yau four-folds with vanishing
electric flux exist. To investigate this further one may well have to
return to the standard approach of analysing the $G$-structure of the most
general ans\"{a}tze for the Killing spinors as in ~\cite{Gauntlett:2004zh}.
However, this will be complicated. 

Another advantage, beyond technical tractability, of the techniques we
have employed in this paper, is that by tracking the $G$-structure
reduction induced by incorporating additional Killing spinors, we have
been able to give a unified treatment of all the wrapped brane and
$AdS$ spacetimes we consider, by deriving the supersymmetry conditions
in every case from the co-associative and associative calibration
conditions. We have also seen how the
R-symmetries of the dual SCFTs are encoded in the supergravity
descriptions of the wrapped brane spacetimes, by elucidating how the
isometries arise in the $AdS$ limits.

It would be interesting to generalise the results here to
cover other wrapped M5-brane geometries. For instance, it should be
straightforward to extend our analysis to the case of M5-branes
wrapping cycles in eight-dimensional manifolds, with $\Spin(7)$,
$\SU(4)$ or $\Symp(2)$ holonomy. In these cases, electric charge can
be induced from the Chern--Simons term in eleven-dimensional
supergravity and so this should require a slight generalisation of the
wrapped-brane ansatz to allow for this flux. The $AdS_3$ gauged
supergravity solutions in~\cite{gkw, gk2} corresponding to M5-branes
wrapping calibrated cycles in eight-dimensional manifolds are of this
type. This generalisation should allow 
for the classification of a variety of $AdS_3$
spacetimes with varying degrees of supersymmetry.

More generally, there are $AdS_{d+1}$ geometries with electric flux which
do not come from wrapped M5-branes, the simplest example being
the basic Freund--Ruben $AdS_4$ solutions which are the near-horizon limit
of a set of M2-branes at the apex of a cone with 
special holonomy contained in $Spin(7)$. 
Furthermore, an interesting example
with dyonic fluxes is that of \cite{warner}, where an
$AdS_4$ solution with both electric and magnetic fluxes is
constructed as the IR fixed point of a supersymmetric flow.
Another natural generalisation, as in the previous paragraph, is thus to
extend the wrapped-brane ansatz to include membrane probes or more
generally dyonic probes which include both membrane and fivebrane
charge. This kind of background appeared in the analysis of the
generic minimally supersymmetric spacetimes with $\bbR^{1,2}$ given
in~\cite{J+D}. 

An auxiliary result of our analysis is that all the supersymmetry
conditions for the wrapped-brane spacetimes could be interpreted in
terms of generalised calibrations and that this gave a relatively simply
way of deriving the constraints on the geometry. A natural conjecture
is that this is a general result. More precisely one might expect the
conditions for supersymmetry for any given background to be equivalent
to a set of eleven-dimensional generalised calibration conditions
related to the allowed set of Killing spinors. In particular, the
analysis of~\cite{gp} implies that when the Killing spinor is
timelike, the eleven-dimensional calibration conditions are indeed
equivalent to the Killing spinor equation. Given an equivalent
statement in the null case, the equivalence of supersymmetry
conditions and the allowed set of generalised calibrations is then
straightforward.

Compared to the success of~\cite{Gauntlett:2004zh} it
has proved difficult to construct new explicit solutions. While we
found some new examples, all were singular. However, in several cases
we have reduced the problem of finding explicit solutions to that of
solving a system of first-order ODEs. One might hope that a more
in-depth (possibly numerical) analysis of these equations 
might lead to new solutions. And, of course, there is
much scope for exploring further generalisations of the gauged
supergravity solutions, which we leave to the future.

\subsection*{Acknowledgements}
OC is supported by EPSRC. DW is supported by the Royal Society through
a University Research Fellowship.


\appendix


\section{Projections and structures}
\label{projs}

In this appendix we list a set of spinor projections which can be used
to define the spinor ans\"{a}tze for the wrapped-brane spacetimes. In all
cases the spinors can be chosen to be eigenspinors of the five
commuting projection operators
\begin{equation}
   \left\{
      \Gamma^{1234}, \Gamma^{3456}, \Gamma^{5678},
      \Gamma^{1357}, \Gamma^{+-}
   \right\} .
\end{equation}
We will be interested in the cases of probe branes wrapping manifolds
with $G_2$, $\SU(3)$ and $\SU(2)$ special holonomy.

\subsubsection*{Co-associative and associative calibrations in $G_2$
  holonomy}

We take the special holonomy geometry
$\bbR^{1,3}\times\mathcal{M}_{G_2}$ with $\bbR^{1,3}$ spanned by
$\{e^+,e^-,e^8,e^9\}$. One can define the four Killing spinors by
\begin{equation}
\label{g2proj}
   \Gamma^{1234}\epsilon^i = \Gamma^{3456}\epsilon^i
     = \Gamma^{1357}\epsilon^i = - \epsilon^i .
\end{equation}
With this definition the $G_2$ structure takes the standard form
\begin{equation}
\label{PhiUp}
\begin{aligned}
   \Phi &= e^{127}+e^{347}+e^{567}+e^{246}-e^{136}-e^{145}-e^{235},\\
   \Upsilon &=
      e^{1234}+e^{3456}+e^{1256}+e^{1357}-e^{1467}-e^{2367}-e^{2457}.
\end{aligned}
\end{equation}

For a probe brane wrapping a co-associative cycle we have $d=1$
in~\eqref{g-ansatz} and the unwrapped world-volume is spanned by
$\{e^+,e^-\}$. We take the brane projection
$\Gamma^{+-1234}\epsilon^i=-\epsilon^i$ or equivalently
\begin{equation}
   \Gamma^{+-}\epsilon^i = \epsilon^i \qquad \text{(co-associative)},
\end{equation}
and $e^8$ and $e^9$ define the overall transverse directions in
$\mathcal{M}_9$.

For a probe brane wrapping an associative cycle we have $d=2$ and the
unwrapped worldvolume is spanned by $\{e^+,e^-,e^8\}$. We take the
brane projection $\Gamma^{+-8567}\epsilon^i=\epsilon^i$ or
equivalently
\begin{equation}
\label{ass-proj}
   \Gamma^{+-}\Gamma^{5678}\epsilon^i = - \epsilon^i
   \qquad \text{(associative)},
\end{equation}
and $e^9$ defines the overall transverse direction in
$\mathcal{M}_8$.

\subsubsection*{K\"ahler and SLAG calibrations in $SU(3)$ holonomy}

The special holonomy geometry is
$\bbR^{1,4}\times\mathcal{M}_{\SU(3)}$ with $\bbR^{1,4}$ spanned by
$\{e^+,e^-,e^7,e^8,e^9\}$. One can define the eight Killing spinors by
\begin{equation}
\label{SU3proj}
   \Gamma^{1234}\epsilon^i = \Gamma^{3456}\epsilon^i = - \epsilon^i .
\end{equation}
The $\SU(3)$ structure then takes the standard form
\begin{equation}
\begin{aligned}
   J &= e^{12} + e^{34} + e^{56} , \\
   \Omega &= (e^1+\ii e^2)\wedge(e^3+\ii e^4)\wedge(e^5+\ii e^6) .
\end{aligned}
\end{equation}
Further projecting under $\Gamma^{1357}$ and comparing
with~\eqref{g2proj} we see that this is equivalent to a pair of $G_2$
structures
\begin{equation}
\label{g2pair}
\begin{aligned}
   \Phi_\pm &= \pm J\wedge e^7- \im\Omega, \\
   \Upsilon_\pm &= \tfrac{1}{2}J\wedge J \pm \re\Omega\wedge e^7.
\end{aligned}
\end{equation}

For a probe brane wrapping a K\"ahler four-cycle we have $d=1$ and the
unwrapped world-volume is spanned by $\{e^+,e^-\}$. We take the brane
projection $\Gamma^{+-1234}\epsilon^i=-\epsilon^i$ or equivalently
\begin{equation}
\label{4in6proj}
   \Gamma^{+-}\epsilon^i = \epsilon^i
   \qquad \text{(K\"ahler four-cycle)},
\end{equation}
and $\{e^7,e^8,e^9\}$ span the overall transverse directions in
$\mathcal{M}_9$.

For a probe brane wrapping a K\"ahler two-cycle we have $d=4$ and the
unwrapped worldvolume is spanned by $\{e^+,e^-,e^7,e^8\}$. We take the
brane projection $\Gamma^{+-7856}\epsilon^i=-\epsilon^i$ or
equivalently
\begin{equation}
\label{2in6proj}
   \Gamma^{+-}\Gamma^{5678}\epsilon^i = - \epsilon^i
   \qquad \text{(K\"ahler two-cycle)},
\end{equation}
and $e^9$ defines the overall transverse direction in $\mathcal{M}_7$.

For a probe brane wrapping a SLAG cycle we have $d=3$ and the
unwrapped worldvolume is spanned by $\{e^+,e^-,e^7\}$. We take the
brane projection $\Gamma^{+-7135}\epsilon^i=\epsilon^i$ or
equivalently
\begin{equation}
\label{slagproj}
   \Gamma^{+-}\Gamma^{1357}\epsilon^i = - \epsilon^i
   \qquad \text{(SLAG cycle)},
\end{equation}
and $e^8$ and $e^9$ define the overall transverse directions in
$\mathcal{M}_8$.

\subsubsection*{K\"ahler calibrations in $SU(2)$ holonomy}

The special holonomy geometry is
$\bbR^{1,6}\times\mathcal{M}_{\SU(2)}$ with $\bbR^{1,6}$ spanned by
$\{e^+,e^-,e^5,\dots,e^9\}$. One can define the 16 Killing
spinors by
\begin{equation}
\label{SU2proj}
   \Gamma^{1234}\epsilon^i = - \epsilon^i .
\end{equation}
The $\SU(2)$ structure then takes the standard form
\begin{equation}\label{A13}
\begin{aligned}
   J^1 &= e^{12}+e^{34}, \\
   J^2 &= e^{14}+e^{23}, \\
   J^3 &= e^{13}-e^{24}.
\end{aligned}
\end{equation}
Further projecting under $\Gamma^{3456}$ and comparing
with~\eqref{SU3proj} we see this is equivalent to a pair of $\SU(3)$
structures
\begin{equation}
\label{SU3pair}
\begin{aligned}
   J &= J^1 \pm e^{56}, \\
   \Omega &= (J^3+\ii J^2)\wedge(e^5\pm\ii e^6) .
\end{aligned}
\end{equation}

For a probe brane wrapping a K\"ahler two-cycle we have $d=4$ and the
unwrapped worldvolume is spanned by $\{e^+,e^-,e^5,e^6\}$. We take the
brane projection $\Gamma^{+-5634}\epsilon^i=-\epsilon^i$ or
equivalently
\begin{equation}
\label{2in4proj}
   \Gamma^{+-}\Gamma^{3456}\epsilon^i = - \epsilon^i
   \qquad \text{(K\"ahler two-cycle)},
\end{equation}
and $\{e^7,e^8,e^9\}$ span the overall transverse direction in
$\mathcal{M}_7$.

\section{AdS limits of wrapped brane metrics}\label{appA}
In this appendix, we will give some further technical discussion of
the assumptions we make in taking the $AdS$ limit of the wrapped-brane
metrics. Specifically, we will show that in the case of one overall
transverse direction, the rotation angle $\theta$ must be independent
of the $AdS$ radial coordinate $r$, so in this case this requirement
need not be imposed as an additional assumption. In the case of two or
three overall transverse directions, we will show that with a suitable
assumption of $r$-independence of the frame rotation, the part of the
$AdS$ radial direction lying in the overall transverse space must in
fact lie entirely along the radial direction of the overall transverse
space, as we assumed in the main text. We will discuss the cases of
one, two, or three overall transverse directions in turn.

\subsection{One overall transverse direction}
There is one overall transverse direction for the cases of branes wrapping
associative three-cycles or K\"{a}hler two-cycles in manifolds with
$SU(3)$ holonomy. Then, necessarily,
$\hat{v}=Ldt$. We want to show that the rotation angle
$\theta$ must be independent of the $AdS$ radial coordinate. We will
see that this follows from the condition that the flux be independent of
the $AdS$ radial coordinate in the $AdS$ limit, together with the fact
that the flux for the wrapped brane metrics is completely determined
by supersymmetry.

We will focus on proving this for the $AdS$ limit of branes wrapping
associative cycles; the argument for K\"{a}hler two-cycles in $SU(3)$
holonomy is
very similar. We have the relationships
\bea
\l^{-1/2}\dd r&=&\sin\theta\hat{u}+\cos\theta\hat{v},\nn
\hat{\r}&=&\cos\theta\hat{u}-\sin\theta\hat{v},\nn
ds^2(\mathcal{M}_{G_2})&=&ds^2(\mathcal{M}_{SU(3)})+\hat{u}\otimes\hat{u},
\eea
where the metric in the $AdS$ limit is
\bea
ds^2=\l^{-1}ds^2(AdS_4)+ds^2(\mathcal{M}_{SU(3)})+\hat{\r}\otimes\hat{\r}.
\eea
By assumption, the metric on $\mathcal{M}_{SU(3)}$ is independent of
$r$. Therefore we may always choose the frame on $\mathcal{M}_{(SU(3)}$ to be
independent of $r$, which implies that $\hat{\r}$ must be independent
of $r$. Now, in the $AdS$ limit, the expression (\ref{bvccc}) for the
flux becomes
\bea
&&\l e^{mr}\dd\Big[\l^{-1}e^{-mr}\Big(\frac{1}{2}J\wedge
J+\mbox{Re}\Omega\wedge[\l^{-1/2}\sin\theta\dd
r+\cos\theta\hat{\r}]\Big)\Big]=\nn&&-(\cos\theta\l^{-1/2}\dd
r-\sin\theta\hat{\r})\wedge F.
\eea
Since $F$ has no components along the $AdS$ radial direction, we may
read it off by comparing the $\dd r$ terms on each side. In
particular, we consider
the components of $F$ on $\mathcal{M}_{SU(3)}$. These are given by
\bea
-\frac{m}{2\cos\theta}J\wedge
J+\frac{\l}{\cos\theta}\tilde{\dd}(\l^{-3/2}\sin\theta\mbox{Re}\Omega),
\eea
where $\tilde{\dd}$ denotes the exterior derivative restricted to
$\mathcal{M}_{SU(3)}$. The coefficient of the $J\wedge J$ part of this
expression is
\bea
\frac{2\sin\theta\l^{-1/2}\mbox{Re}\mathcal{W}_1-m}{2\cos\theta},
\eea
where we have used $\tilde{\dd}\Omega=\mathcal{W}_1J\wedge J+\dots$. Since $\l$ and
$\mathcal{W}_1$ are independent of $r$, this flux component is
independent of $r$ iff $\theta$ is independent of $r$, as claimed. By
a very similar argument, one may show that $\theta$ must also be
independent of $r$ for K\"{a}hler two-cycles in $SU(3)$ holonomy manifolds.

\subsection{Two overall transverse directions}
Now we turn to the case of two overall transverse directions. We will use
a slightly different set-up to that of the main text. We define the
``wrapped brane frame''
\bea
e^1&=&Ldy^1,\nn
e^2&=&Ldy^2,\nn
e^3&=&\hat{u},
\eea
where $y^{1,2}$ are cartesian coordinates on the overall transverse
space. We define the ``$AdS$ frame''
\bea
(e^1)^{\prime}&=&\l^{-1/2}dr,
\eea
with $(e^2)^{\prime}$, $(e^3)^{\prime}$ given by
\bea
(e^A)^{\prime}=R_{AB}e^B,
\eea
for some $Spin(3)$ matrix $R$, with $A,B=1,2,3$. Here we are viewing the $AdS$
radial direction as arising from $e^{1,2,3}$ through what is {\it a
  priori} a completely
general frame rotation. We wish to show that under the assumption that
the matrix $R$ is
independent of the $AdS$ radial coordinate $r$, we may always choose
it such that
\bea
(e^2)^{\prime}&=&\hat{\r},\nn
(e^3)^{\prime}&=&\frac{1}{2m}\l\r\dd\phi,
\eea
with $\hat{\r}$ as given in section 5. This is equivalent to the
statement that assuming $r$-independence of $R$, the part of the $AdS$
radial direction lying in the overall transverse space lies entirely
along the radial direction of the overall transverse space.

In general, we have
\bea
dy^1&=&\lambda^{-1}e^{-2mr}(R_{11}\lambda^{-1/2}dr+R_{21}(e^2)^{\prime}+R_{31}(e^3)^{\prime}),\\
dy^2&=&\lambda^{-1}\ee^{-2mr}(R_{12}\lambda^{-1/2}dr+R_{22}(e^2)^{\prime}+R_{32}(e^3)^{\prime}).
\eea
Now, given that $R$ is independent of $r$, demanding that $dy^{1,2}$
are closed, and using the fact that $R$ is a special
orthogonal matrix, we find the following expressions for $(e^2)^{\prime}$,
$(e^3)^{\prime}$:
\bea
(e^2)^{\prime}&=&-\frac{\lambda}{2mR_{13}}(R_{32}d(\lambda^{-3/2}R_{11})-R_{31}d(\lambda^{-3/2}R_{12})),\\
(e^3)^{\prime}&=&\frac{\lambda}{2mR_{13}}(R_{22}d(\lambda^{-3/2}R_{11})-R_{21}d(\lambda^{-3/2}R_{12})).
\eea
Next, defining coordinates $\rho,\phi$ such that
\bea
\lambda^{-3/2}R_{11}&=&\rho\cos\phi,\\
\lambda^{-3/2}R_{12}&=&\rho\sin\phi,
\eea
the $(e^2)^{\prime}$, $(e^3)^{\prime}$ become
\bea
(e^2)^{\prime}&=&\frac{1}{2m\lambda^{1/2}}\Big(\frac{R_{23}}{\rho
    R_{13}}d\rho-R_{33}d\phi\Big),\\
(e^3)^{\prime}&=&\frac{1}{2m\lambda^{1/2}}\Big(\frac{R_{33}}{\rho
    R_{13}}d\rho+R_{23}d\phi\Big).
\eea
We still have the freedom to perform rotations about the $AdS$ radial
direction, which we may exploit to choose a simpler frame. Thus, we
perform a $Spin(2)$ rotation in the $2^{\prime}3^{\prime}$ plane,
according to
\bea
(\hat{e}^2)^{\prime}&=&\frac{1}{\sqrt{1-R_{13}^2}}\Big(R_{23}(e^2)^{\prime}+R_{33}(e^3)^{\prime}\Big)=\frac{\lambda
d\rho}{2m\sqrt{1-\lambda^3\rho^2}}=\hat{\r},\nn
(\hat{e}^3)^{\prime}&=&\frac{1}{\sqrt{1-R_{13}^2}}\Big(-R_{33}(e^2)^{\prime}+R_{23}(e^3)^{\prime}\Big)=\frac{1}{2m}\lambda\rho
d\phi,
\eea
and so obtain the desired result.

\subsection{Three overall transverse directions}
The analysis with three overall transverse directions is qualitatively
very similar to that with two, though it is technically somewhat more
involved. We now take our ``wrapped brane frame'' to be given by
\bea
e^1&=&Ldy^1,\nn
e^2&=&Ldy^2,\nn
e^3&=&Ldy^3,\nn
e^4&=&\hat{u}.
\eea
Our ``$AdS$ frame'' is given by
\bea
(e^A)^{\prime}=R_{AB}e^B,
\eea
where now $A,B=1,...,4$, $R$ is a $Spin(4)$ matrix and
$(e^{1})^{\prime}=\l^{-1/2}dr$. As before, we want to show that
assuming that $R$ is independent of the $AdS$ radial coordinate $r$,
we may always choose it such that
\bea
(e^2)^{\prime}&=&\hat{\r},\nn
(e^3)^{\prime}&=&\frac{1}{2m}\l\r\dd \chi,\nn
(e^4)^{\prime}&=&\frac{1}{2m}\l\r\sin\chi\dd\phi,
\eea
and that therefore the part of the $AdS$ radial direction lying in the
overall transverse space must lie entirely along the radial direction
of the overall transverse space.

In general, we have
\bea
dy^1&=&\lambda^{-1}e^{-2mr}(R_{11}\lambda^{-1/2}dr+R_{21}(e^2)^{\prime}+R_{31}(e^3)^{\prime}+R_{41}(e^4)^{\prime}),\\
dy^2&=&\lambda^{-1}\ee^{-2mr}(R_{12}\lambda^{-1/2}dr+R_{22}(e^2)^{\prime}+R_{32}(e^3)^{\prime}+R_{42}(e^{4})^{\prime}),\\
dy^3&=&\lambda^{-1}\ee^{-2mr}(R_{13}\lambda^{-1/2}dr+R_{23}(e^2)^{\prime}+R_{33}(e^3)^{\prime}+R_{43}(e^{4})^{\prime}).
\eea
Now, given that $R$ is independent of $r$, demanding that $dy^{1,2,3}$
are closed, we get
\bea
(e^2)^{\prime}&=&\frac{\l}{2mR_{14}}\e^{ijk4}R_{3i}R_{4j}d(\l^{-3/2}R_{1k}),\\
(e^3)^{\prime}&=&\frac{\l}{2mR_{14}}\e^{ijk4}R_{4i}R_{2j}d(\l^{-3/2}R_{1k}),\\
(e^4)^{\prime}&=&\frac{\l}{2mR_{14}}\e^{ijk4}R_{2i}R_{3j}d(\l^{-3/2}R_{1k}),
\eea
where $i,j,k=1,...,4$ and $\e^{1234}=1$. We still have the freedom to
perform $Spin(3)$ rotations about the $AdS$ radial direction, to
simplify the frame. To this end, we define coordinates $\r,\chi,\phi$
such that
\bea
\l^{-3/2}R_{11}&=&\r\sin\chi\sin\phi,\\
\l^{-3/2}R_{12}&=&\r\sin\chi\cos\phi,\\
\l^{-3/2}R_{13}&=&\r\cos\chi.
\eea
With this choice of coordinates, our frame is given by
\bea
\begin{pmatrix}(e^2)^{\prime}\\(e^3)^{\prime}\\(e^4)^{\prime}\end{pmatrix}=\frac{Q}{2m}\begin{pmatrix}\frac{\l}{\sss}d\r\\\l\r
  d\chi\\\l\r\sin\chi d\phi\end{pmatrix},
\eea
where the matrix $Q$ is given by
\bea
Q=\frac{1}{\l^{3/2}\r}\begin{pmatrix}R_{24}&\frac{1}{\l^{3/2}\r\sin\chi
    R_{14}}(R_{24}R_{13}-\l^3\r^2\e^{34ij}R_{3i}R_{4j})&\frac{1}{\sin\chi}\e^{12ij}R_{3i}R_{4j}\\R_{34}&\frac{1}{\l^{3/2}\r\sin\chi
    R_{14}}(R_{34}R_{13}-\l^3\r^2\e^{34ij}R_{4i}R_{2j})&\frac{1}{\sin\chi}\e^{12ij}R_{4i}R_{2j}\\R_{44}&\frac{1}{\l^{3/2}\r\sin\chi
    R_{14}}(R_{44}R_{13}-\l^3\r^2\e^{34ij}R_{2i}R_{3j})&\frac{1}{\sin\chi}\e^{12ij}R_{2i}R_{3j}\end{pmatrix}.
\eea
It may be verified that $Q$ is an element of $Spin(3)$. Therefore we
may rotate about the $AdS$ radial direction to get a new frame, given
by
\bea
\begin{pmatrix}(\hat{e}^2)^{\prime}\\(\hat{e}^3)^{\prime}\\(\hat{e}^4)^{\prime}\end{pmatrix}=Q^{-1}\begin{pmatrix}(e^2)^{\prime}\\(e^3)^{\prime}\\(e^4)^{\prime}\end{pmatrix}=\frac{1}{2m}\begin{pmatrix}\frac{\l}{\sss}d\r\\\l\r
  d\chi\\\l\r\sin\chi d\phi\end{pmatrix},
\eea
as required.

\section{Sample calculations of the supersymmetry conditions}
In this appendix, we will give more details of a representative
example of the derivation of the $AdS$ supersymmetry conditions from
the wrapped brane supersymmetry conditions. We will focus on the
derivation of the $\mathcal{N}=2$ $AdS_4$ supersymmetry conditions from the SLAG
supersymmetry conditions. We have the following expressions for the
basis one-forms in the ``wrapped brane frame'' in terms of the
coordinates in the ``$AdS$ frame'':
\bea
\hat{u}&=&\l^{-1/2}\sss dr+\frac{\l^{5/2}\r d\r}{2m\sss},\\
Ldt&=&\l\r dr-\frac{\l}{2m}d\r,\\
Ltd\phi&=&-\frac{\l\r}{2m}d\phi.
\eea
The $SU(3)$ forms appearing in (\ref{219})-(\ref{222}) decompose into $SU(2)$ forms according to
\bea
J&=&J^1+\hat{w}\wedge \hat{u},\nn
\mbox{Re}\Omega&=&J^3\wedge\hat{w}-J^2\wedge\hat{u},\nn
\mbox{Im}\Omega&=&J^2\wedge\hat{w}+J^3\wedge\hat{u},
\eea
where the $J^i$ are given by (\ref{A13}). We define the new frame
\bea
\hat{\rho}&=&\frac{\l}{2m\sss} d\r,\\
\hat{\phi}&=&\frac{\l\r}{2m}d\phi,\\
\hat{r}&=&\l^{-1/2}dr.
\eea
The frame in the directions transverse to the $AdS$
factor is independent of $r$. Equation (\ref{221}) becomes
\bea
d\Big[\l^{-1/2}\ee^{-mr}\Big(J^1+\hat{w}\wedge(\l^{-1/2}\sss
dr+\frac{\l^{5/2}\r}{2m\sss}d\r)\Big)\Big]=0,
\eea
which reads
\bea\label{poi}
d[\l^{-1}\sss \hat{w}]=m(\l^{-1/2}J^1+\l\r \hat{w}\wedge\hat{\r}),
\eea
which is (\ref{101}). Next, imposing $\hat{r}\lrcorner F=0$, we
find that (\ref{222}) becomes
\bea\label{poiu}
d[\l^{-3/2}J^3\wedge \hat{w}-\r J^2\wedge \hat{\r}]&=&0,\\ \label{flx}
\hat{r}\wedge\l^2\Big[d[\l^{-2}\sss J^2]-3m(\l^{-3/2}J^3\wedge
\hat{w}-\r J^2\wedge\hat{\r})\Big]&=&\star_8F.
\eea
The first of these equations is (\ref{1011}). Next, (\ref{220}) becomes
\bea\label{poiuy}
d\phi\wedge d[J^2\wedge \hat{w}+\frac{1}{\l^{3/2}\r}J^3\wedge
\hat{\r}]=0.
\eea
This is consistent with (\ref{10111}) but does not imply it. However,
observe that (\ref{poi}) implies that
\bea
\partial_{\phi}(\l^{-1}\sss \hat{w})=0,
\eea
and the exterior derivative of (\ref{poi}) implies that
\bea
\partial_{\phi}(\l^{-1/2}J^1)=\partial_{\phi}\Big(\frac{\l^2}{\sss}\hat{w}\Big)=0,
\eea
and therefore that
\bea\label{c15}
\partial_{\phi}\l=\partial_{\phi}\hat{w}=\partial_{\phi}J^1=0.
\eea
Then (\ref{poiu}) implies that
\bea\label{c16}
\partial_{\phi}J^2=\partial_{\phi}J^3=0,
\eea
and hence (\ref{poi}), (\ref{poiu}) and (\ref{poiuy}) imply
equation (\ref{10111}). Furthermore, (\ref{c15}) and (\ref{c16}) imply
that $\partial_{\phi}$ is Killing, as claimed in the main text.

The final SLAG torsion condition is $\mbox{Re}\Omega \wedge
d\mbox{Re}\Omega=0$. Rewriting this as
\bea
\l^{-3/2}\mbox{Re}\Omega \wedge
d(\l^{-3/2}\mbox{Re}\Omega)=0,
\eea
and using (\ref{poiu}), we get
\bea
&&(\l^{-3/2}J^3\wedge \hat{w}-\r J^2\hat{\r}\wedge d(\l^{-2}\sss
J^2)=\frac{1}{2m\r}d(\l^{-1}J^2\wedge J^2)\wedge
d\r\nn&&=\frac{1}{2m\r}d(\l^{-1}J^1\wedge J^1)\wedge d\r=0.
\eea
But this is automatically satisfied as a consequence of the exterior
derivative of (\ref{poi}),
which states that
\bea\label{A35}
d(\l^{-1/2}J^1)=-\frac{3\r}{1-\l^3\r^2}d\l\wedge\hat{w}\wedge\hat{\r}-\frac{m\l^{3/2}\r}{\sss}J^1\wedge \hat{\r}.
\eea

It remains to obtain the expression (\ref{515}) for the flux. To do
this, we use (\ref{rrrr}), with $v=\hat{v}=Ldt$. We get
\bea
&&\hat{\phi}\wedge d\Big[J^2\wedge\hat{w}+J^3\wedge\Big(\l^{-1/2}\sss
dr+\frac{\l^{5/2}\r}{2m\sss}d\r\Big)\Big]=\nn&&(\l\r dr-\frac{\l}{2m}d\r)\wedge
F.
\eea
Since $F$ has no components along the $AdS$ radial direction, we may
simply read it off directly by comparing the $dr$ terms on both sides.
It is simple to see that the Killing vector $\partial_\phi$ leaves the
flux invariant.

\section{Deriving the LLM conditions}\label{app2}
In this appendix, we will show that the general solution of our $\mathcal{N}=2$ $AdS_5$
supersymmetry conditions precisely satisfies the conditions derived by LLM~\cite{LLM}. To
begin, let us define
\bea
e^1&=&\frac{\l}{2m\sss}\tilde{e}^1,\\
e^2&=&\frac{\l}{2m\sss}\tilde{e}^2,\\
e^3&=&\frac{\sss}{m\l^{1/2}}\tilde{e}^3.
\eea
Then equations (\ref{65}) become
\bea\label{7114}
d\tilde{e}^1&=&-\frac{\l^3\r}{2(1-\l^3\r^2)}d\r\wedge\tilde{e}^1+\tilde{e}^{23},\\\label{7115}
d\tilde{e}^2&=&-\frac{\l^3\r}{2(1-\l^3\r^2)}d\r\wedge\tilde{e}^2+\tilde{e}^{31},\\\label{7116}
2d\tilde{e}^3&=&-\frac{\l^3}{(1-\l^3\r^2)^2}\tilde{e}^{12}-\frac{\r}{(1-\l^3\r^2)^2}\Big(\partial_{\r}\l^3\tilde{e}^{12}\nn&-&[\tilde{\partial}_2\l^3\tilde{e}^1-\tilde{\partial}_1\l^3\tilde{e}^2]\wedge
d\r\Big).
\eea
Equation (\ref{7114}) implies that we may write
\bea
\tilde{e}^1=\ee^{\frac{1}{2}D(\r,x^a)}\hat{e}^1(x^a),
\eea
where the $x^a$, $a=1,2,3$ are some coordinates on the three-space
spanned by the $\tilde{e}^a$ (which we refer to as the base), and furthermore that
\bea
\l^3=-\frac{\partial_{\r}D}{\r(1-\r\partial_{\r}D)}.
\eea
Similarly from (\ref{7115}), we find that we may write
\bea
\tilde{e}^2=\ee^{\frac{1}{2}D}\hat{e}^2(x^a).
\eea
Then if we denote the exterior derivative restricted to the base by
$\tilde{d}$, the remaining content of (\ref{7114}), (\ref{7115}) is
\bea\label{7120}
\tilde{d}\hat{e}^1&=&-\frac{1}{2}\tilde{d}D\wedge\hat{e}^1+\hat{e}^2\wedge\tilde{e}^3,\\\label{7121}
\tilde{d}\hat{e}^2&=&-\frac{1}{2}\tilde{d}D\wedge\hat{e}^2-\hat{e}^1\wedge\tilde{e}^3.
\eea
Next, from (\ref{7116}), we find that
\bea
(\partial_{\r}\tilde{e}^3)_3=0.
\eea
Therefore we may choose our coordinates such that as a vector
\bea
\tilde{e}^3=\frac{\partial}{\partial x^3},
\eea
and as a one-form
\bea
\tilde{e}^3=(dx^3+V_{\hat{i}}(\r,x^a)\hat{e}^i),
\eea
where $\hat{i}=1,2$. Now, by taking the $\r$ derivative of
(\ref{7120}) and (\ref{7121}), we find that
\bea\label{7125}
\partial_{\r x^3}D&=&0,\\\label{7126}
\partial_{\r}(\hat{\partial}_2D-2V_{\hat{1}})&=&0,\\\label{7127}
\partial_{\r}(\hat{\partial}_1D+2V_{\hat{2}})&=&0.
\eea
We are free to shift the definition of $D$ by an arbitrary function of
the $x^a$. Thus (\ref{7125}) implies that we may always take
\bea
D=D(\r,x^1,x^2).
\eea
Then the $x^3$ dependence of $\hat{e}^1$, $\hat{e}^2$ is fixed by
(\ref{7120}) and (\ref{7121}) to be given by
\bea
\partial_{x^3}\hat{e}^1&=&-\hat{e}^2,\\
\partial_{x^3}\hat{e}^2&=&\hat{e}^1.
\eea
Therefore, we have
\bea
\hat{e}^1&=&\sin x^3\overline{e}^1(x^1,x^2)+\cos
x^3\overline{e}^2(x^1,x^2),\\
\hat{e}^2&=&-\cos x^3\overline{e}^1(x^1,x^2)+\sin
x^3\overline{e}^2(x^1,x^2).
\eea
The absence of a term with $\tilde{e}^3$ on the RHS of (\ref{7116})
implies that
\bea
V_{\hat{i}}=V_{\hat{i}}(\r,x^1,x^2).
\eea
Returning to equations (\ref{7126}) and (\ref{7127}), and denoting the
exterior derivative restricted to the two-space spanned by the
$\overline{e}^i$ by $\overline{d}$,  we
see that we may write
\bea
V=\frac{1}{2}\star_2\overline{d}D+\xi(x^a),
\eea
for some one-form $\xi$. Since $V$ and $D$ are independent of $x^3$,
then so also is $\xi$. We still have two gauge degrees of freedom
left, which we may use to set $\xi=0$. To see this, observe that we may shift $D$ by an
arbitrary function of $x^i$, $D\rightarrow D+2f(x^1,x^2)$, and by means
of a shift in $x^3$, we may set $V\rightarrow V+dg$, for some
arbitrary function $g(x^1,x^2)$. Thus we may always take $\xi=0$ if we
can solve
\bea
dg=\star_2df+\xi.
\eea
Taking the exterior derivative of this equation and its dual, we find
that we may set $\xi=0$ if we can find functions $f,g$ that solve
\bea
d\star_2df&=&-d\xi,\nn
d\star_2dg&=&d\star_2\xi.
\eea
But these are just two independent copies of Poisson's equation in two Riemannian
dimensions, and we may always find a solution of each in a local
coordinate patch. Therefore we may always take
$V=\frac{1}{2}\star_2\overline{d}D$, and (\ref{7120}) and (\ref{7121}) reduce to
\bea
d\overline{e}^i=0,
\eea
for which we take the local solution
\bea
\overline{e}^i=dx^i.
\eea
It may now be verified that upon inserting all the conditions we have
derived above, equation (\ref{7116}) reduces to the Toda equation
\bea
(\partial_{x^1}^2+\partial_{x^2}^2)D+\partial_{\r}^2\ee^D=0.
\eea
Given a solution of this equation, the metric is given by
\bea
ds^2&=&\frac{1}{\l
  m^2}\Big[ds^2(AdS_5)+\frac{\l^3}{4}\Big(\frac{1}{1-\l^3\r^3}(d\r^2+\ee^Ddx^idx^i)+\r^2ds^2(S^2)\Big)\nn&&+(1-\l^3\r^2)(dx^3+V_idx^i)^2\Big],
\eea
where
\bea
\l^3&=&-\frac{\partial_{\r}D}{\r(1-\r\partial_{\r}D)},\\
V&=&\frac{1}{2}\star_2\overline{d}D,
\eea
and the flux may be read off from (\ref{6.9}). As claimed in the main text, these are precisely the
LLM conditions, which are given in \cite{LLM} for $m=1/2$.

\end{document}